\documentclass[prd,nofootinbib,twocolumn]{revtex4}

\usepackage{amsmath,amssymb}
\usepackage{rotating}
\usepackage{mathrsfs}
\usepackage{pstricks}
\usepackage{pst-func}
\usepackage{graphicx}
\usepackage{refcount}
\usepackage{cases}
\usepackage{hyperref}

\allowdisplaybreaks

\begin{document}

\title{Bell's Theorem Versus Local Realism in a Quaternionic Model of Physical Space}

\author{Joy Christian}

\email{jjc@alum.bu.edu}

\affiliation{Einstein Centre for Local-Realistic Physics, 15 Thackley End, Oxford OX2 6LB, United Kingdom}

\begin{abstract}
In the context of EPR-Bohm type experiments and spin detections confined to spacelike hypersurfaces, a local, deterministic and realistic model within a Friedmann-Robertson-Walker spacetime with a constant spatial curvature (${S^3}$) is presented that describes simultaneous measurements of the spins of two fermions emerging in a singlet state from the decay of a spinless boson. Exact agreement with the probabilistic predictions of quantum theory is achieved in the model without data rejection, remote contextuality, superdeterminism or backward causation. A singularity-free Clifford-algebraic representation of ${S^3}$ with vanishing spatial curvature and non-vanishing torsion is then employed to transform the model in a more elegant form. Several event-by-event numerical simulations of the model are presented, which confirm our analytical results with the accuracy of 4 parts in ${10^4}$. Possible implications of our results for practical applications such as quantum security protocols and quantum computing are briefly discussed. 
\end{abstract}

\maketitle

\tighten

\renewcommand\bf{\mathbf}

\section{Introduction}\label{sec:introduction}

According to the statistical ensemble interpretation of quantum theory, the wave function, which is obtained as a solution of Schr\"odinger's equation, does not provide a complete description of an individual physical system but rather of an  ensemble  of  similar  physical systems. Because it removes many conceptual difficulties, this interpretation of quantum theory was preferred by Einstein \cite{Ballentine}. In particular, within this interpretation the wave function can be  interpreted epistemically, merely as a compendium of our knowledge of the physical systems. This knowledge then changes upon a measurement of the system, not unlike how our knowledge of the state of a classical coin changes in a coin-toss experiment once the coin lands on its head or on its tail. Consequently, the mystery surrounding the so-called collapse of the wave function and related paradoxes are removed. This would be the end of all conceptual difficulties except for two obstacles:

(1) The first of these two obstacles is Bell's theorem \cite{Bell-1964}, which suggests that the quantum probabilities encoded in the wave function are not epistemic but, in fact, objective, and the randomness observed in the quantum world is not reducible to the coin-toss randomness observed in classical physics. 

(2) The second obstacle is the following question: If the quantum probabilities describe merely a compendium of our knowledge of the physical systems, then how come that knowledge, encoded in the amplitude of those probabilities called the wave function, evolves objectively, under a specific dynamical equation called the Schr\"odinger's equation?

Our goal in this paper is to remove obstacle (1) for making the statistical interpretation of quantum mechanics plausible. Removing obstacle (2) may then become viable in the light of the success of our strategy of removing obstacle (1). 

Now, it is well known that, unlike our most fundamental theories of space and time, quantum theory is incompatible with local causality, if we take Realism for granted, as usually done. This fact was recognized in 1935 by Einstein, Podolsky, and Rosen (EPR) \cite{EPR}. They hoped, however, that perhaps quantum mechanics can be completed into a locally causal theory by addition of supplementary (or ``hidden") parameters. Today such hopes of maintaining both locality and realism within physics seem to have been undermined by Bell's theorem \cite{Bell-1964}, with considerable support from experiments \cite{Clauser,Aspect,Weihs,Hensen,Giustina,Shalm,Rauch}. Bell set out to prove that no physical theory that is realistic as well as local in the sense espoused by Einstein can reproduce all of the statistical predictions of quantum mechanics. By contrast, in this paper we present a physically well-motivated constructive counterexample to Bell's theorem by deriving the strong singlet correlations using the powerful language of Geometric Algebra \cite{Clifford,Dorst}.

The novelty of our manifestly geometrical derivation of the singlet correlations is likely to be of considerable interest, not only in the investigations of the foundations of quantum mechanics, but also in their technological applications in quantum cryptography, quantum computing, and quantum security protocols \cite{Milburn}. Indeed, the Geometric Algebra techniques we have used in this paper are already being employed in numerous engineering and technological applications \cite{Dorst,Application-1,Application-2,Application-3,Application-4,Hitzer}. They range from applications in computer vision technology to those in aviation engineering.

The value of the Geometric Algebra techniques in computer vision modeling and aviation engineering is not surprising. One of the central concerns in these disciplines is how to represent rotations in the physical space in a singularity-free manner. If one tries to represent rotations using Euler angles, for example, then one runs into a gimbal lock problem when two of the three rotation axes align themselves, because one of the rotation degrees of freedom is then lost. This is a fatal problem for airplane controls, especially if an airplane is in a steep ascent or descent. But within Geometric Algebra one can represent rotations {\it smoothly} using the even subalgebra of the Clifford algebra ${Cl_{3,0}}$, as we have done in this paper. As a result, the gimbal lock problem can be avoided entirely.

In addition to this, the locally causal understanding of quantum correlations we have offered in this paper may also assist in the quantum technological applications, such as in quantum cryptography, quantum security protocols, and quantum computing. As noted by Zeilinger in his recent survey of such technologies \cite{Zeilinger}, it is often not easy to foresee how fundamental results can turn out to have practical applications too. As he recalls, rather surprisingly, experimental work following the purely philosophical questions concerning local realism did, in time, become an important ingredient in a number of applications in quantum information technology, including in the invention of quantum repeaters that may connect future quantum computers with each other at long distances\cite{Zeilinger}. Another application inspired by fundamental work is entanglement-based quantum cryptography. Entanglement swapping has also been applied in the so-called loophole-free tests of Bell's theorem. Such experiments suggest that unconditionally secure quantum cryptography is possible, since quantum cryptography based on the traditional interpretation of Bell's theorem can provide unconditional security. An eavesdropper can no longer avoid detection in an experiment that correctly follows the protocol.

As we will discuss in Section \ref{limit}, the traditional interpretation of Bell's theorem is recovered from our model in the ${S^3\rightarrow{\mathrm{I\!R}^3}}$ limit. The question then is: How are the practical applications mentioned above affected when the geometrical and topological properties of the quaternionic 3-sphere are taken into account? The key features of these properties are the M\"obius-like twists in the Hopf bundle of ${S^3}$, and (if ${S^3}$ is taken as a physical space) the related conservation of the spin angular momentum captured in Eq.~(\ref{596}) below. As we shall see, when these properties of ${S^3}$ are taken into account correctly, the strong correlations are easily reproduced. The open question then is: How are the technological applications such as quantum computing, quantum cryptography, and quantum security protocols affected when the geometrical and topological properties of the quaternionic 3-sphere are taken into account? The quantitative  results presented in this paper are likely to have serious implications for this question.

\begin{figure*}[t]
\hrule
\scalebox{1}{
\begin{pspicture}(1.2,-2.5)(4.2,2.5)

\psline[linewidth=0.1mm,dotsize=3pt 4]{*-}(-2.51,0)(-2.5,0)

\psline[linewidth=0.1mm,dotsize=3pt 4]{*-}(7.2,0)(7.15,0)

\psline[linewidth=0.4mm,arrowinset=0.3,arrowsize=3pt 3,arrowlength=2]{->}(-2.5,0)(-3,1)

\psline[linewidth=0.4mm,arrowinset=0.3,arrowsize=3pt 3,arrowlength=2]{->}(-2.5,0)(-3,-1)

\psline[linewidth=0.4mm,arrowinset=0.3,arrowsize=3pt 3,arrowlength=2]{->}(7.2,0)(8.3,0.5)

\psline[linewidth=0.4mm,arrowinset=0.3,arrowsize=3pt 3,arrowlength=2]{->}(7.2,0)(7.4,1.3)

\psline[linewidth=0.4mm,arrowinset=0.3,arrowsize=2pt 3,arrowlength=2]{->}(4.2,0)(4.2,1.1)

\psline[linewidth=0.4mm,arrowinset=0.3,arrowsize=2pt 3,arrowlength=2]{->}(0.5,0)(0.5,1.1)

\pscurve[linewidth=0.2mm,arrowinset=0.2,arrowsize=2pt 2,arrowlength=2]{->}(4.0,0.63)(3.85,0.45)(4.6,0.5)(4.35,0.65)

\put(4.1,1.25){{\large ${{\bf s}_2}$}}

\pscurve[linewidth=0.2mm,arrowinset=0.2,arrowsize=2pt 2,arrowlength=2]{<-}(0.35,0.65)(0.1,0.47)(0.86,0.47)(0.75,0.65)

\put(0.4,1.25){{\large ${{\bf s}_1}$}}

\put(-2.4,+0.45){{\large ${\bf 1}$}}

\put(6.8,+0.45){{\large ${\bf 2}$}}

\put(-3.35,1.35){{\large ${\bf a}$}}

\put(-3.5,-1.7){{\large ${\bf a'}$}}

\put(8.5,0.52){{\large ${\bf b}$}}

\put(7.3,1.5){{\large ${\bf b'}$}}

\put(1.8,-0.65){\large source}

\put(0.99,-1.2){\large ${\pi^0\longrightarrow\,e^{-}+\,e^{+}\,}$}

\put(1.11,0.5){\large total spin = 0}

\psline[linewidth=0.3mm,linestyle=dashed](-2.47,0)(2.1,0)

\psline[linewidth=0.4mm,arrowinset=0.3,arrowsize=3pt 3,arrowlength=2]{->}(-0.3,0)(-0.4,0)

\psline[linewidth=0.3mm,linestyle=dashed](2.6,0)(7.2,0)

\psline[linewidth=0.4mm,arrowinset=0.3,arrowsize=3pt 3,arrowlength=2]{->}(5.0,0)(5.1,0)

\psline[linewidth=0.1mm,dotsize=5pt 4]{*-}(2.35,0)(2.4,0)

\pscircle[linewidth=0.3mm,linestyle=dashed](7.2,0){1.3}

\psellipse[linewidth=0.2mm,linestyle=dashed](7.2,0)(1.28,0.3)

\pscircle[linewidth=0.3mm,linestyle=dashed](-2.51,0){1.3}

\psellipse[linewidth=0.2mm,linestyle=dashed](-2.51,0)(1.28,0.3)

\end{pspicture}}
\hrule
\caption{A spin-less neutral pion decays into an electron-positron pair (such a photon-less decay is quite rare but not impossible, and will suffice for our theoretical purposes here \cite{RSOS}). Measurements of spin components on each separated fermion are performed at remote stations ${\mathbf{1}}$ and ${\mathbf{2}}$, providing binary outcomes along arbitrary directions such as ${\mathbf a}$ and ${\mathbf b}$. The conservation of angular momentum dictates that the net spin of the pair remains zero [cf. Eq.~(\ref{56})].}
\vspace{0.2cm}
\hrule
\label{fig1}
\end{figure*}

\section{An Oversight in the proof of Bell's theorem}\label{secII}

Returning to Bell's theorem, despite its ambitious scope, its proof is mathematically rather simple \cite{Bell-1964}. It is based on mathematical inequalities discovered by Boole over one hundred years before Bell \cite{Boole}, and proceeds as follows:

Consider the EPR \cite{EPR} type spin-${\frac{1}{2}}$ experiment, as originally proposed by Bohm \cite{Bohm} (cf. Fig.~\ref{fig1}). Alice is free to choose a detector direction ${\mathbf{a}}$ or ${\mathbf{a'}}$ and Bob is free to choose a detector direction ${\mathbf{b}}$ or ${\mathbf{b'}}$ to detect spins of the fermions they receive from a common source, at a space-like distance from each other. The objects of interest then are the bounds on the sum of possible averages put together in the manner of Clauser, Horne, Shimony, and Holt (CHSH) \cite{CHSH,Clauser}.
\begin{equation}
{\cal E}({\mathbf{a}},\,{\mathbf{b}})\,+\,{\cal E}({\mathbf{a}},\,{\mathbf{b'}})\,+\,{\cal E}({\mathbf{a'}},\,{\mathbf{b}})\,-\,{\cal E}({\mathbf{a'}},\,{\mathbf{b'}})\,, \label{B1-11-2}
\end{equation}
with each average of a product of local functions defined as
\begin{align}
{\cal E}({\mathbf{a}},\,{\mathbf{b}})\,&=\lim_{\,n\,\gg\,1}\left[\frac{1}{n}\sum_{k\,=\,1}^{n}\,
{\mathscr A}({\mathbf{a}},\,{\lambda}^k)\;{\mathscr B}({\mathbf{b}},\,{\lambda}^k)\right] \notag \\
&\equiv\,\Bigl\langle\,{\mathscr A}_{k}({\mathbf{a}})\,{\mathscr B}_{k}({\mathbf{b}})\,\Bigr\rangle\,,\label{expect-2}
\end{align}
where ${\mathscr A({\mathbf{a}},\,{\lambda}^k)\equiv {\mathscr A}_{k}({\mathbf{a}})=\pm1}$ and ${\mathscr B({\mathbf{b}},\,{\lambda}^k)\equiv {\mathscr B}_{k}({\mathbf{b}})}$ ${=}$ ${\pm1}$ are the respective {\it local} measurement results of Alice and Bob, with ${\lambda^k}$ being a hidden variable for the ${k^{th}}$ run of the experiment. Now, since ${{\mathscr A}_{k}({\mathbf{a}})=\pm1}$ and ${{\mathscr B}_{k}({\mathbf{b}})=\pm1}$, the average of their product is ${-1\leqslant\Bigl\langle\,{\mathscr A}_{k}({\mathbf{a}})\,{\mathscr B}_{k}({\mathbf{b}})\,\Bigr\rangle\leqslant +1}$. As a result, we can immediately read off the upper and lower bounds on the sequence of four averages considered in (\ref{B1-11-2}):
\begin{align}
-\,4\,\leqslant&\,\Bigl\langle\,{\mathscr A}_{k}({\mathbf{a}})\,{\mathscr B}_{k}({\mathbf{b}})\,\Bigr\rangle\,+\, \Bigl\langle\,{\mathscr A}_{k}({\mathbf{a}})\,{\mathscr B}_{k}({\mathbf{b'}})\,\Bigr\rangle\, \notag \\
&+\,\Bigl\langle\,{\mathscr A}_{k}({\mathbf{a'}})\,{\mathscr B}_{k}({\mathbf{b}})\,\Bigr\rangle\,-\, \Bigl\langle\,{\mathscr A}_{k}({\mathbf{a'}})\,{\mathscr B}_{k}({\mathbf{b'}})\,\Bigr\rangle\,\leqslant\,+\,4\,. \label{3-1}
\end{align}
Next, using the rule ``a sum of averages is equal to the average of the sum'', we replace the above sum of four {\it separate} averages of ${\pm1}$ numbers with the single average of their sum:
\begin{align}
{\cal E}(&{\mathbf{a}},\,{\mathbf{b}})\,+\,{\cal E}({\mathbf{a}},\,{\mathbf{b'}})\,+\,{\cal E}({\mathbf{a'}},\,{\mathbf{b}})\,-\,{\cal E}({\mathbf{a'}},\,{\mathbf{b'}}) \notag \\
&\longrightarrow\Bigl\langle{\mathscr A}_{k}({\mathbf{a}}){\mathscr B}_{k}({\mathbf{b}})+{\mathscr A}_{k}({\mathbf{a}}){\mathscr B}_{k}({\mathbf{b'}}) \notag \\
&\;\;\;\;\;\;\;\;\;\;\;\;\;\;\;+{\mathscr A}_{k}({\mathbf{a'}}){\mathscr B}_{k}({\mathbf{b}})-{\mathscr A}_{k}({\mathbf{a'}}){\mathscr B}_{k}({\mathbf{b'}})\Bigr\rangle. \label{rep}
\end{align}
This step allows us to reduce the sum (\ref{3-1}) of four averages to
\begin{equation}
\Bigl\langle{\mathscr A}_{k}({\mathbf{a}})\big\{{\mathscr B}_{k}({\mathbf{b}})+{\mathscr B}_{k}({\mathbf{b'}})\big\}\,+\,{\mathscr A}_{k}({\mathbf{a'}})\big\{{\mathscr B}_{k}({\mathbf{b}})-{\mathscr B}_{k}({\mathbf{b'}})\big\}\Bigr\rangle.\label{absurd}
\end{equation}
And because ${{\mathscr B}_{k}({\mathbf{b}})=\pm1}$, if ${|{\mathscr B}_{k}({\mathbf{b}})+{\mathscr B}_{k}({\mathbf{b'}})|=2}$, then ${|{\mathscr B}_{k}({\mathbf{b}})-{\mathscr B}_{k}({\mathbf{b'}})|=0}$, and vice versa. Consequently, using ${{\mathscr A}_{k}({\mathbf{a}})=\pm1}$, it is easy to conclude that the absolute value of the above average cannot exceed 2, just as Bell concluded~\cite{Bell-1964}: 
\begin{align}
-\,2\,\leqslant\,\Bigl\langle\,&{\mathscr A}_{k}({\mathbf{a}})\,{\mathscr B}_{k}({\mathbf{b}})\,+\,{\mathscr A}_{k}({\mathbf{a}})\,{\mathscr B}_{k}({\mathbf{b'}}) \notag \\
&+\,{\mathscr A}_{k}({\mathbf{a'}})\,{\mathscr B}_{k}({\mathbf{b}})\,-\,{\mathscr A}_{k}({\mathbf{a'}})\,{\mathscr B}_{k}({\mathbf{b'}})\,\Bigr\rangle\,\leqslant\,+\,2\,.\label{5-1}
\end{align}
On the other hand, because the expectation value analogous to (\ref{expect-2}) predicted by quantum mechanics for the singlet state is
\begin{equation}
{\cal E}_{QM}({\mathbf{a}},\,{\mathbf{b}})=\Bigl\langle\,{\mathscr A}_{k}({\mathbf{a}})\,{\mathscr B}_{k}({\mathbf{b}})\,\Bigr\rangle=-\,{\mathbf{a}}\cdot{\mathbf{b}}=-\cos(\mathbf{a}, \mathbf{b}) \label{expect-qm}
\end{equation}
(cf. Ref.~\cite{Bohm}), quantum mechanical predictions exceed the bounds of ${\pm\,2}$ on the inequalities (\ref{5-1}) for some combinations of angles among the experimental directions such as ${\mathbf{a}}$ and ${\mathbf{b}}$:  
\begin{equation}
-2\sqrt{2}\leqslant{\cal E}({\mathbf{a}},\,{\mathbf{b}})+{\cal E}({\mathbf{a}},\,{\mathbf{b'}})+{\cal E}({\mathbf{a'}},\,{\mathbf{b}})-{\cal E}({\mathbf{a'}},\,{\mathbf{b'}})\leqslant 2\sqrt{2}.
\end{equation}
Consequently, Bell concluded that no physical theory that is both realistic and local in the senses envisaged by Einstein can reproduce all of the predictions of quantum mechanics.

However, the above formal proof of Bell's theorem does not justify its physically radical conclusion. As innocuous as the step (\ref{rep}) in the proof may seem mathematically, it is, in fact, an illegitimate step physically, because what is being averaged on its right-hand are {\it unobservable} and {\it unphysical} quantities. Indeed, the pairs of measurement directions ${({\mathbf{a}},\,{\mathbf{b}})}$, ${({\mathbf{a}},\,{\mathbf{b'}})}$, ${({\mathbf{a'}},\,{\mathbf{b}})}$, and ${({\mathbf{a'}},\,{\mathbf{b'}})}$ are {\it mutually exclusive measurement directions}, corresponding to {\it incompatible} experiments which cannot be performed simultaneously. Each pair can be used by Alice and Bob for a given experiment, for all runs, but no two of the four pairs can be used by them simultaneously. This is because Alice and Bob do not have the ability to make measurements along counterfactually possible pairs of directions such as ${({\mathbf{a}},\,{\mathbf{b}})}$ and ${({\mathbf{a}},\,{\mathbf{b'}})}$ simultaneously. Alice, for example, can make measurements along ${\mathbf{a}}$ or ${\mathbf{a'}}$, but not along ${\mathbf{a}}$ {\it and} ${\mathbf{a'}}$ at the same time \cite{Bell-oversight}.

Thus, contrary to the claim of Bell's theorem, it is not the objectively measurable predictions of quantum mechanics that rule out the possibility of a local and realistic theory. It is the {\it ad hoc} and unjustified assumption of three or four physically incompatible experiments, any one of which might be performed on a given occasion, but only one of which can, in fact, be performed in practice, and in reality \cite{Bell-oversight,RSOS}.

We are therefore justified in ignoring the physical claim of Bell's theorem in this paper, which then opens up the opportunity of a {\it constructive} approach to reproducing the quantum mechanical correlations in any locally causal manner\footnote{Incidentally, criticisms of Bell's theorem began soon after the publication of Bell's paper in 1964 \cite{Bell-1964}. They included criticisms from critics such as Louis de Broglie \cite{deBroglie} and Max Jammer\cite{Jammer}, and have never ceased ever since. During the intermediate years of its acceptance by the larger physics community, one of the ardent critics of Bell's theorem has been Arthur Fine, who pointed out in the 1980s that Bell's argument depends on considering joint probability distribution of three or four mutually incompatible observables, which is not a justifiable assumption \cite{Fine,Fine-2}. By now there exists a vast literature on various criticisms of Bell's theorem. While not all such criticisms are reliable or of high quality, there do exist a number of high quality criticisms, published in respectable peer-review journals, such as by Karl Hess \cite{Hess}, Willem M. de Muynck \cite{Muynck}, Itamar Pitowsky \cite{Pitowsky}, Hans de Raedt \cite{deRaedt}, and Andrei Khrennikov \cite{Khrennikov}, to mention just a few of many.}. In what follows, we demonstrate that it is, in fact, possible to reproduce the statistical predictions of quantum states such as the EPR-Bohm state \cite{Bohm} in a strictly locally causal manner in a Friedmann-Robertson-Walker spacetime, albeit viewed as a {\it non}-cosmological solution of Einstein's field equations of general relativity. To demonstrate this, we shall follow the locally causal framework proposed by Bell himself \cite{Bell-1964} (which is reviewed in the Appendix below for convenience), using the powerful language of Geometric Algebra \cite{Clifford,Dorst}.

\section{Two Particles Entangled in a Singlet State}

As noted, a locally causal description of the measurement of the spins of two spacelike separated spin-${\frac{1}{2}}$ particles that are products of the decay of a single spin-zero particle has been considered by Bell in his pioneering paper \cite{Bell-1964}. Based on Bohm's version of the EPR thought experiment, he considered a pair of spin-${\frac{1}{2}}$ particles, moving freely after the decay in opposite directions, with particles ${1}$ and ${2}$ subject (respectively) to spin measurements along independently chosen unit directions ${\mathbf{a}}$ and ${\mathbf{b}}$, which may be located at a spacelike distance from one another. If initially the emerging pair has vanishing total spin,
then its quantum mechanical spin state can be described by the entangled singlet state, 
\begin{equation}
|\Psi_{\mathbf n}\rangle=\frac{1}{\sqrt{2}}\Bigl\{|{\mathbf n},\,+\rangle_1\otimes
|{\mathbf n},\,-\rangle_2\,-\,|{\mathbf n},\,-\rangle_1\otimes|{\mathbf n},\,+\rangle_2\Bigr\}\,,
\label{single}
\end{equation}
with ${\mathbf n}$ as arbitrary direction and ${{\boldsymbol\sigma}\cdot{\mathbf n}\,|{\mathbf n},\,\pm\rangle\,=\,\pm\,|{\mathbf n},\,\pm\rangle}$ describing the quantum mechanical eigenstates in which the particles have spin up or down in the units of ${\hbar=2}$. Here ${\boldsymbol\sigma}$ is the spin ``vector'' ${(\sigma_x,\,\sigma_y,\,\sigma_z)}$ composed of Pauli matrices.

Our interest lies in an event-by-event reproduction of the probabilistic predictions of this entangled quantum state in a locally causal manner \cite{Bell-1964}. For any freely chosen measurement directions ${\mathbf{a}}$ and ${\mathbf{b}}$ in space there would be nine possible outcomes of the experiment in general, regardless of the distance between the directions. If we denote the angle between ${\mathbf{a}}$ and ${\mathbf{b}}$ by ${\eta_{{\mathbf{a}}{\mathbf{b}}}}$ and the local measurement results ${0}$, ${+1}$, or ${-1}$ about these directions by ${\mathscr A}$ and ${\mathscr B}$, then quantum mechanics is well known to predict the following joint probabilities for these results:
\begin{align}
P_{12}^{+-}(\eta_{{\mathbf{a}}{\mathbf{b}}})&=P\{{\mathscr A}=+1,\;{\mathscr B}=-1
\;|\;\eta_{{\mathbf{a}}{\mathbf{b}}}\} \notag \\
&=\frac{1}{2}\cos^2\left(\frac{\eta_{{\mathbf{a}}{\mathbf{b}}}}{2}\right)\!,\\
P_{12}^{++}(\eta_{{\mathbf{a}}{\mathbf{b}}})&=P\{{\mathscr A}=+1,\;{\mathscr B}=+1
\;|\;\eta_{{\mathbf{a}}{\mathbf{b}}}\} \notag \\
&=\frac{1}{2}\sin^2\left(\frac{\eta_{{\mathbf{a}}{\mathbf{b}}}}{2}\right)\!, \\
P_{12}^{-+}(\eta_{{\mathbf{a}}{\mathbf{b}}})&=P_{12}^{+-}(\eta_{{\mathbf{a}}{\mathbf{b}}}), \\
P_{12}^{--}(\eta_{{\mathbf{a}}{\mathbf{b}}})&=P_{12}^{++}(\eta_{{\mathbf{a}}{\mathbf{b}}}), \\
P_{12}^{+0}(\eta_{{\mathbf{a}}{\mathbf{b}}})&=P_{12}^{-0}(\eta_{{\mathbf{a}}{\mathbf{b}}})=P_{12}^{0+}(\eta_{{\mathbf{a}}{\mathbf{b}}})=P_{12}^{0-}(\eta_{{\mathbf{a}}{\mathbf{b}}})=0,
\end{align}
and
\begin{equation}
P_{12}^{00}(\eta_{{\mathbf{a}}{\mathbf{b}}})=0,
\end{equation}
where the superscript 0 indicates no detection and the subscripts 1 and 2 label the particles \cite{Pearle}. The probability that the spin of particle 1 will be detected parallel to ${\mathbf{a}}$ (regardless of whether particle 2 itself is detected) is also predicted by quantum mechanics. It is given by
\begin{equation}
P_1^{+}({\mathbf{a}})=P_1^{-}({\mathbf{a}})=\frac{1}{2},
\end{equation}
and likewise for particle 2 being detected parallel to ${\mathbf{b}}$. In what follows our goal is to demonstrate that, at least in the Friedmann-Robertson-Walker spacetime ${{\mathrm{I\!R}}\times \Sigma}$ with a constant spatial curvature, the above probabilities can be reproduced within the original local model of Bell \cite{Bell-1964}.

\section{Within the Spatial slice of a  Friedmann-Robertson-Walker spacetime}

\begin{figure*}[t]
\hrule
\begin{pspicture}(-6.25,-3.5)(2.5,2.4)

\put(-9.75,1.62){{t}}

\put(-8.07,0.57){{${S^3}$}}

\psline[linewidth=0.2mm,arrowinset=0.3,arrowsize=2pt 4,arrowlength=2]{->}(-9.47,0.67)(-9.47,1.87)

\psline[linewidth=0.2mm,arrowinset=0.3,arrowsize=2pt 4,arrowlength=2]{->}(-9.47,0.67)(-8.17,0.67)

\put(-5.4,1.0){{${{\mathscr A}=\pm1}$}}

\put(-4.9,0.2){{${\bf a}$}}

\put(-2.37,-2.0){{${({\bf e}_o,\,{\bf s}_o)}$}}

\put(-1.88,-2.5){{${\lambda}$}}

\put(-3.05,-2.9){${Is_1(\lambda)}$ \;\;\;\;\;\;${Is_2(\lambda)}$}

\put(0.53,1.0){{${{\mathscr B}=\pm1}$}}

\put(1.03,0.2){{${\bf b}$}}

\pscircle[fillcolor=black,fillstyle=solid](-4.8,0.67){0.05}

\pscircle[fillcolor=black,fillstyle=solid](1.13,0.67){0.05}

\pscircle[linewidth=0.3mm,linestyle=dashed](-4.8,0.77){1.035}

\pscircle[linewidth=0.3mm,linestyle=dashed](1.13,0.77){1.035}

\psline[linewidth=0.2mm,linestyle=dashed]{-}(-3.9,1.27)(0.07,-3.0)

\psline[linewidth=0.2mm,linestyle=dashed]{-}(-5.7,1.27)(-9.67,-3.0)

\psline[linewidth=0.2mm,linestyle=dashed]{-}(2.03,1.27)(6.0,-3.0)

\psline[linewidth=0.2mm,linestyle=dashed]{-}(0.23,1.27)(-3.74,-3.0)

\end{pspicture}
\hrule
\caption{The local results ${{\mathscr A}({\bf a};\,{\bf e}_o,\,{\bf s}_o)}$ and ${{\mathscr B}({\bf b};\,{\bf e}_o,\,{\bf s}_o)}$ are deterministically brought about by the initial state ${({\bf e}_o,\,{\bf s}_o)}$ originating in the overlap of the two backward light cones of Alice and Bob (cf. Figure 6.4 of Ref.~\cite{Bell-1990}). In the Clifford-algebraic representation of the local model the initial state is a possible orientation (or handedness) ${\lambda}$ of the 3-sphere, and the constituent spins are represented by ${Is_1(\lambda)}$ and ${Is_2(\lambda)}$ (cf. Section \ref{GA}). The black dot in the overlap of the backward light cones represents the source (or creation event) of the constituent spins, and the black dots in the spacelike separated stations represent the detection events.}
\vspace{0.25cm}
\label{fig2}
\hrule
\end{figure*}

Friedmann-Robertson-Walker (FRW) spacetimes are a set of solutions of Einstein's field equations of general relativity. They are widely used in cosmology to study the spacetime geometries governing the dynamics of our universe \cite{d'Inverno}. In our view, they are therefore the most appropriate spacetimes within which all Bell-test experiments should be discussed. It is generally accepted that the geometries of our universe are described by the Friedmann-Robertson-Walker line element
\begin{equation}
ds^2=dt^2-a^2(t)\,d\Sigma^2, \;\;\,d\Sigma^2=\left[\frac{d\rho^2}{1-\kappa\,\rho^2}+\rho^2 d\Omega^2\right]\!, \label{frw}
\end{equation}
where ${a(t)}$ is the scale factor, ${\Sigma}$ is a spacelike hypersurface, ${\rho}$ is the radial coordinate within ${\Sigma}$, ${\kappa}$ is the ``normalized'' curvature of ${\Sigma}$, and ${\Omega}$ is a solid angle within ${\Sigma}$ \cite{d'Inverno}. Since we are primarily concerned with a galactic, solar, or terrestrial scenario, in what follows, without loss of generality, we will restrict our attention to the current epoch of the cosmos by setting the scale factor ${a(t) = 1}$ in the above line element. 

Now it is well known that the rescaled (or ``normalized'') curvature ${\kappa}$ can take only three possible values, $+1$, $-1$, or $0$. For ${\kappa = 0}$ the above FRW line element describes ordinary flat space, or ${\mathrm{I\!R}^3}$, in spherical coordinates. For ${\kappa = +1}$ it describes the metric of a 3-sphere, ${S^3}$, with constant positive curvature. And for ${\kappa = -1}$ it describes a three-dimensional hyperboloid in four-dimensional Minkowski space (${AdS_3}$), with constant negative curvature. Thus the only metric giving a closed and compact space without running off to infinity is ${S^3}$, for ${\kappa = +1}$. The geometry and topology of this space is therefore disciplined enough to give rise to the observed strong correlations of the singlet state. To verify this, in what follows we consider a spacelike hypersurface ${\Sigma=S^3}$ in the FRW spacetime, which can be recovered from the line element (\ref{frw}) by introducing a new coordinate ${\chi\!=\sin^{-1}\!\rho}$ (for a detailed derivation see Section 22.8 of Ref.~\cite{d'Inverno}).  

Now the tangent bundle of ${S^3}$ happens to be trivial ({\it i.e.}, it happens to be a product space): ${{\mathrm{T}}S^3 = S^3 \times{\mathrm{I\!R}}^3}$. This renders the tangent space at each point of ${S^3}$ to be isomorphic to ${{\mathrm{I\!R}}^3}$. Thus local experiences of the experimenters within ${S^3}$ are no different from those of their counterparts within ${{\mathrm{I\!R}}^3}$. The global topology of ${S^3}$, however, is dramatically different from that of ${{\mathrm{I\!R}}^3}$. In particular, the triviality of ${{\mathrm{T}}S^3}$ means that ${S^3}$ is parallelizable \cite{Nakahara}. Therefore, a global {\it anholonomic} frame can be specified on ${S^3}$ that fixes each of its points uniquely \cite{Nakahara,Christian}. Such a frame renders ${S^3}$ diffeomorphic to the symmetry group ${\mathrm{SU}(2)}$ --- {\it i.e.}, to the set of all unit quaternions \cite{Meister}:
\begin{equation}
S^3\,=\,\left\{\,{\mathbf{H}}(I\cdot{\mathbf{v}},\,\eta)\,\;\bigg|\;\,||\,{\mathbf{H}}(I\cdot{\mathbf{v}},\,\eta)\,||\,=\,1\,\right\}\!. \label{onpara}
\end{equation}
Here we have parameterized each quaternion ${{\mathbf{H}}\in S^3}$ as
\begin{equation}
{\mathbf{H}}(I\cdot{\mathbf{v}},\,\eta)=\exp{\left\{\,(I\cdot{\mathbf{v}})\,\eta\,\right\}} \label{offpara}
\end{equation}
such that ${I\cdot{\mathbf{v}}}$ is a bivector rotating about some vector ${{\mathbf{v}}\in{\mathrm{I\!R}}^3}$, and ${\eta}$ is half of the angle by which ${\mathbf{H}}$ stands rotated about ${\mathbf{v}}$. Up to sign, ${I\cdot{\bf v}}$ is identical to the dual of ${\bf v}$, with the unit and oriented trivector ${I:={\bf e}_x{\bf e}_y{\bf e}_z}$ as a {\it pseudoscalar}, because it commutes with all other elements of the Clifford algebra ${Cl_{3,0}}$. The trivector can therefore be used to represent a volume form on a given manifold, such as the quaternionic 3-sphere we are considering. As in these definitions, in what follows we will be using the notation of Geometric Algebra \cite{Christian,Dorst,Clifford}. Accordingly, all vector fields in ${{\mathrm{I\!R}}^3}$ such as ${\mathbf{v}}$ and ${\mathbf{w}}$ will be assumed to satisfy the geometric product
\begin{equation}
{\mathbf v}\,{\mathbf w}\,=\,{\mathbf v}\cdot{\mathbf w}\,+\,{\mathbf v}\wedge{\mathbf w},
\end{equation}
with the duality relation ${{\mathbf v}\wedge{\mathbf w}=I\cdot({\mathbf v}\times{\mathbf w})}$. In the next steps it will be useful to recall that ${({\mathbf v}\wedge{\mathbf w})^{\dagger}=-({\mathbf v}\wedge{\mathbf w})}$.

In what follows, we will not be using the time coordinate appearing in (\ref{frw}) explicitly. Instead, we will follow the practice of defining the measurement events in terms of the initial and final instants of time as usually done within the context of Bell's local-realistc model \cite{Bell-1964,Clauser}. Readers who are not familiar with the above practice are urged to review the Appendix A below before proceeding further. We also postpone the full spacetime covariant investigation of local causality until Section \ref{GA}, where we show how and why only the spacelike components of the relativistic spins enter the calculations of EPR correlations even though the two spins themselves emerge relativistically from the overlap of the backward light cones of Alice and Bob.

Consider now two unit quaternions from the closed set ${S^3}$, say ${{\mathbf{P}}_o({\mathbf{n}}\wedge{\mathbf{e}}_o,\,\eta_{{\mathbf{n}}{\mathbf{e}}_o})}$ and ${{\mathbf{Q}}_o({\mathbf{z}}\wedge{\mathbf{s}}_o,\,\eta_{{\mathbf{z}}{\mathbf{s}}_o})}$, defined as
\begin{equation}
{\mathbf{P}}_o\,=\,\cos(\,\eta_{{\mathbf{n}}{\mathbf{e}}_o})\,+\,\frac{{\mathbf{n}}\wedge{\mathbf{e}}_o}{||{\mathbf{n}}\wedge{\mathbf{e}}_o||}\,\sin(\,\eta_{{\mathbf{n}}{\mathbf{e}}_o})\label{3}
\end{equation}
and
\begin{equation}
{\mathbf{Q}}_o\,=\,\cos(\,\eta_{{\mathbf{z}}{\mathbf{s}}_o})\,+\,\frac{{\mathbf{z}}\wedge{\mathbf{s}}_o}{||{\mathbf{z}}\wedge{\mathbf{s}}_o||}\,\sin(\,\eta_{{\mathbf{z}}{\mathbf{s}}_o}), \label{Q}
\end{equation}
where ${{\mathbf n}\in T_pS^3 \cong {\mathrm{I\!R^3}}}$ is an arbitrary unit vector in the tangent space ${T_pS^3}$ at some point ${p}$ of ${S^3}$, ${\mathbf{z}}$ is a fixed reference vector in ${T_qS^3}$ at a different point ${q}$ of ${S^3}$, and ${{\mathbf{e}}_o}$ and ${{\mathbf{s}}_o}$ are two other tangential vectors in ${T_qS^3}$. Here the bivector ${I\cdot{\mathbf{e}}_o}$ may be thought of as representing an individual spin within the pair of decaying particles in the singlet state, and the bivector ${I\cdot{\mathbf{s}}_o}$ may be thought of as representing the spin of the composite pair \cite{Pearle}. Note that, although ${{\mathbf{P}}_o}$ and ${{\mathbf{Q}}_o}$ are normalized to unity, their sum ${{\mathbf{P}}_o+{\mathbf{Q}}_o}$ need not be. In fact, they satisfy the following triangle inequality for arbitrary pairs of such quaternions,
\begin{equation}
||{\mathbf{P}}_o+\,{\mathbf{Q}}_o|| \,\leqslant\, ||{\mathbf{P}}_o||\,+\,||{\mathbf{Q}}_o||\,,\label{5}
\end{equation}
reflecting the metrical structure of ${S^3\!}$. Moreover, since ${S^3}$ is closed under multiplication, we also have ${||{\mathbf{P}}_o{\mathbf{Q}}_o||=1}$.

Within the freedom provided by this inequality, the above constraints lead us to the following choice for the set of initial (or {\it complete} \cite{Bell-1964}) states ${\left({\mathbf{P}}_o,\,{\mathbf{Q}}_o\right)}$ of our physical system:
\begin{equation}
\Lambda=\left\{\left({\mathbf{P}}_o,\,{\mathbf{Q}}_o\right)\;\Big|\;||{\mathbf{P}}_o+{\mathbf{Q}}_o||\,=\,{\cal N}(\,\eta_{{\mathbf{n}}{\mathbf{e}}_o},\;\eta_{{\mathbf{z}}{\mathbf{s}}_o})\;\;\forall\,{\mathbf{n}}\,\right\}\!,\label{7}
\end{equation}
with the value ${\cal N}$ of the norm given by the {\it ansatz} 
\begin{equation}
{\cal N}(\,\eta_{{\mathbf{n}}{\mathbf{e}}_o},\;\eta_{{\mathbf{z}}{\mathbf{s}}_o})\!=\!1+\sin^2(\eta_{{\mathbf{n}}{\mathbf{e}}_o})+\!\left[-1+\frac{2}{\sqrt{1+3\left(\frac{\eta_{{\mathbf{z}}{\mathbf{s}}_o}}{\kappa\pi}\right)}}\!\right]^2\!\!\!\!, \label{D}
\end{equation}
which is necessarily a function of the variable angles ${\eta_{{\mathbf{n}}{\mathbf{e}}_o}}$ and ${\eta_{{\mathbf{z}}{\mathbf{s}}_o}}$. It represents a local-realistic counterpart of the singlet state (\ref{single}). It would be different, for example, for a mixed state\footnote{The analysis presented in this paper is generalizable to mixed states. However, the ansatz for the norm ${{\cal N}(\,\eta_{{\mathbf{n}}{\mathbf{e}}_o},\;\eta_{{\mathbf{z}}{\mathbf{s}}_o})}$ appearing in Eq.~(\ref{D}) would be different for a mixed state [or any quantum state other than (\ref{single})], depending on the detailed characteristics of the mixed state. It is beyond the scope and purpose of this paper to investigate the question of mixed states further. The goal of the paper is to investigate local causality for the pure singlet state (\ref{single}) within the context of Bell's theorem, because that is the quantum state considered in the proof of Bell's theorem \cite{Bell-1964}. It is not the goal of the present paper to reproduce all quantum mechanical predictions for any quantum state. Moreover, unlike the pure entangled state (\ref{single}), any mixture of product states will not yield correlations as strong as (\ref{single}) does (i.e., correlations will not deviate much from linear correlations). Therefore mixed states are of little interest in the context of Bell's theorem.}, or for any quantum state other than (\ref{single}). Note also that we have allowed all three possible curvatures of ${\Sigma}$, with ${\kappa=-1}$ being equivalent to ${\eta_{{\bf z}{\bf s}_o}\!\rightarrow 2\pi-\eta_{{\bf z}{\bf s}_o}}$. The significance of this form of ${\cal N}$ will become clear soon. 

If we now substitute expression (\ref{D}) into the inequality
\begin{equation}
||{\bf P}_o||^2\,\geqslant\,||{\bf P}_o+\,{\bf Q}_o||\,-\,1\,,\label{6}
\end{equation}
which follows from multiplying the inequality (\ref{5}) with ${||{\bf P}_o||=1}$ on both sides and simplifying, then [upon using
\begin{equation}|
|{\bf P}_o||^2\,=\,\,\cos^2(\,\eta_{{\bf n}{\bf e}_o})\,+\,\sin^2(\,\eta_{{\bf n}{\bf e}_o}) 
\end{equation}
from Eq.~(\ref{3})] the triangle inequality (\ref{5}) simplifies to
\begin{equation}
|\cos(\,\eta_{{\bf n}{\bf e}_o})| \,\geqslant\,
-1\,+\,\frac{2}{\sqrt{1+3\left(\frac{\eta_{{\bf z}{\bf s}_o}}{\kappa\pi}\right)}\,}. \label{C}
\end{equation}
In what follows it is very important to recognize that this constraint is simply an expression of the intrinsic metrical and topological structures of ${S^3}$, and as such it holds for {\it all} vectors ${\bf n}$ for a given pair of initial states ${({\bf e}_o,\,{\bf s}_o)}$; and, conversely, for {\it all} pairs of initial states ${({\bf e}_o,\,{\bf s}_o)}$ for a given choice of vector ${\bf n}$. This can be easily verified by starting, for example, with a different pair of quaternions, say with the pair ${{\bf P}_o'({\bf n}'\wedge{\bf e}_o,\,\eta_{{\bf n}'{\bf e}_o})}$ and ${{\bf Q}_o({\bf z}\wedge{\bf s}_o,\,\eta_{{\bf z}{\bf s}_o})}$, where
\begin{equation}
{\bf P}_o'\,=\,\cos(\,\eta_{{\bf n}'{\bf e}_o})\,+\,\frac{{\bf n}'\wedge{\bf e}_o}{||{\bf n}'\wedge{\bf e}_o||}\,\sin(\,\eta_{{\bf n}'{\bf e}_o}), \label{3y}
\end{equation}
and arriving at a similar constraint as the one in Eq.~(\ref{C}):
\begin{equation}
|\cos(\,\eta_{{\bf n}'{\bf e}_o})| \,\geqslant\,
-1\,+\,\frac{2}{\sqrt{1+3\left(\frac{\eta_{{\bf z}{\bf s}_o}}{\kappa\pi}\right)}\,}. \label{Cy}
\end{equation}
This procedure can then be repeated for {\it all} vectors ${{\bf n}'}$, and---for a given vector ${\bf n}$---for {\it all} pairs of states ${({\bf e}_o',\,{\bf s}_o')}$.

If we now let ${{\bf e}_o\in T_qS^3}$ and ${{\bf s}_o\in T_qS^3}$ be two random vectors, uniformly distributed over ${S^2}$, and let ${\eta_{{\bf z}{\bf s}_o}}$ be a random scalar, uniformly distributed over ${[0,\,\pi]}$, then we can simplify the set (\ref{7}) of complete or initial states as
\begin{equation}
\Lambda\!=\!\left\{\!({\bf P}_o,\,{\bf Q}_o)\,\Bigg|\,
|\cos(\eta_{{\bf n}{\bf e}_o})| \geqslant -1+\frac{2}{\sqrt{1+3\left(\frac{\eta_{{\bf z}{\bf s_o}}}{\kappa\pi}\right)}}\,\forall\,{\bf n}\!\right\}\!\!. \label{lamset}
\end{equation}
By the previous results this set is invariant under the rotations of ${\bf n}$. Consequently, we identify ${\bf n}$ as a detector direction, and define the measurement events observed by (say) Alice and Bob---along their {\it freely chosen} detector directions ${{\bf n}={\bf a}}$ and ${{\bf n}={\bf b}}$---by two functions of the form
\begin{equation}
\pm\,1\,=\,{\mathscr A}({\bf a};\,{\bf e}_o,\,{\bf s}_o)\!:
{\mathrm{I\!R}}^3\!\times\Lambda\longrightarrow S^3\cong {\mathrm{SU}}(2) \label{nomap-1-2jim}
\end{equation}
and
\begin{equation}
\pm\,1\,=\,{\mathscr B}({\bf b};\,{\bf e}_o,\,{\bf s}_o)\!:
{\mathrm{I\!R}}^3\!\times\Lambda\longrightarrow S^3\cong {\mathrm{SU}}(2). \label{nomap-1-2job}
\end{equation}

These functions are identical to those considered by Bell \cite{Bell-1964} apart from the choice of their codomain, which is now the compact space ${S^3}$ instead of a subset of ${\mathrm{I\!R}}$. That such maps indeed exist can be seen easily by noting that ${{\bf P}_o\rightarrow\pm1}$ as ${\eta_{{\bf n}{\bf e}_o}\rightarrow 0}$ or ${\pi}$. More explicitly, we construct
\begin{align}
S^3\ni\pm 1&={\mathscr A}({\bf a};\,{\bf e}_o,\,{\bf s}_o) \notag \\ 
&=-\,
{\mathrm{sign}}\{\cos(\eta_{{\bf a}{\bf e}_o})\}\,\;\text{for a given}\;{\bf s}_o \label{aamset}
\end{align}
and
\begin{align}
S^3\ni\pm 1&={\mathscr B}({\bf b};\,{\bf e}_o,\,{\bf s}_o) \notag \\
&=+\,
{\mathrm{sign}}\{\cos(\eta_{{\bf b}{\bf e}_o})\}\,\;\text{for the same}\;{\bf s}_o.\label{bamset}
\end{align}
Evidently, these functions define strictly local, realistic, and deterministically determined measurement events. Apart from the common cause ${({\bf e}_o,\,{\bf s}_o)}$, which originates in the overlap of the backward light cones of Alice and Bob as shown in Fig.~\ref{fig2}, the event ${{\mathscr A}=\pm1}$ depends {\it only} on the measurement direction ${\bf a}$ chosen freely by Alice; and analogously, apart from the common cause ${({\bf e}_o,\,{\bf s}_o)}$, the event ${{\mathscr B}=\pm1}$ depends {\it only} on the measurement direction ${\bf b}$ chosen freely by Bob. In particular, the function ${{\mathscr A}({\bf a};\,{\bf e}_o,\,{\bf s}_o)}$ {\it does not} depend on either ${\bf b}$ or ${\mathscr B}$, and the function ${{\mathscr B}({\bf b};\,{\bf e}_o,\,{\bf s}_o)}$ {\it does not} depend on either ${\bf a}$ or ${\mathscr A}$, just as demanded by Bell's formulation of local causality \cite{Bell-1964}.

\section{Calculation of Joint Probability for Observing Measurement Events}

Now, to calculate the joint probabilities for observing the events ${{\mathscr A}=\pm1}$ and ${{\mathscr B}=\pm1}$ simultaneously along the directions ${\bf a}$ and ${\bf b}$, we follow the well known analysis carried out by Pearle for a formally similar local model \cite{Pearle}. Pearle begins by representing each pair of decaying particles by a point ${\bf r}$ in a state space made out of a ball of unit radius in ${\mathrm{I\!R^3}}$. His state space is thus a well known representation of the group SO(3), each point of which corresponding to a rotation, with the direction ${\bf r}$ of length ${0\leqslant r\leqslant 1}$ from the origin representing the axis of rotation and the angle ${{\pi}r}$ representing the angle of rotation. The identity rotation corresponds to the point at the center of the ball. If we now identify the boundaries of two such unit balls, then we recover our 3-sphere, diffeomorphic to the double covering group of SO(3), namely ${\mathrm{SU}(2)}$. The pair of particles in this state space is represented by the quaternion ${{\bf Q}_o}$ defined in Eq.${\,}$(\ref{Q}), which is rotating about the axis ${\frac{{\bf z}\times{\bf s}_o}{||\,{\bf z}\times{\bf s}_o||}}$ by the angle ${2\eta_{{\bf z}{\bf s}_o}}$, with the unit vector ${\bf s_o}$ sweeping a 2-sphere within the 3-sphere \cite{Christian,disproof}.

The relationship between the rotation angle ${{\pi}r}$ within Pearle's state space SO(3) and the rotation angle ${2\eta_{{\bf z}{\bf s}_o}}$ within our state space ${{\mathrm{SU}(2)}\cong S^3}$ turns out to be simple:
\begin{numcases}{\!\!\cos\left(\frac{\pi}{2}r\right)\!=\!}
\!-1+\frac{2}{\sqrt{1+3\left(\frac{\eta_{{\bf z}{\bf s}_o}}{\kappa\pi}\right)}\,}\,=\, f(\eta_{{\bf z}{\bf s}_o}), &  \label{f} \\
\!-1+\frac{2}{\sqrt{4-3\left(\frac{\eta_{{\bf z}{\bf s}_o}}{\kappa\pi}\right)}\,}\,=\,f(\pi-\eta_{{\bf z}{\bf s}_o}). & \label{fp}
\end{numcases}
This can be recognized by first solving Eq.${\,}$(\ref{f}) for ${\frac{\eta_{{\bf z}{\bf s}_o}}{\kappa\pi}}$ and then differentiating the solution with respect to ${r}$, which gives the probability density worked out by Pearle: 
\begin{equation}
p(r) = \frac{1}{\kappa\pi}\frac{\;d\eta_{{\bf z}{\bf s}_o}}{dr}(r)=\frac{4\pi}{3} \frac{\sin(\frac{\pi}{2}r)}{\left\{1 + \cos\left(\frac{\pi}{2} r\right)\right\}^ 3}\,,\;\;\;0\leqslant r\leqslant 1. \label{pdf}
\end{equation}
This function specifies the distribution of probability that a pair of particles is represented by the point ${\bf r}$ in the unit ball. Integrating this distribution from ${0}$ to ${r}$ we may also obtain the cumulative probability distribution in the ball:
\begin{equation}
C(r)\,=\int_0^rp(u)\;du\,=\,-\frac{1}{3} + \frac{4}{3\left\{1+\cos\left(\frac{\pi}{2}r\right)\right\}^2}\,. \label{30}
\end{equation}
This function specifies the probability of finding the pair in any state up to the state ${\bf r}$ within Pearle's state space. From solving Eq.${\,}$(\ref{f}) we see, however, that it is equal to our ratio ${\frac{\eta_{{\bf z}{\bf s}_o}}{\kappa\pi}}$, and therefore also specifies the probability of finding the pair in any initial state up to the state ${{\bf s}_o}$.

For a given reference vector ${\bf z}$, the above relations allow us to translate between our representation in terms of the states ${({\bf e}_o,\,{\bf s}_o)}$ in ${\mathrm{SU}(2)}$ and Pearle's representation in terms of the states ${\bf r}$ in SO(3). We can therefore rewrite our geometrical constraint (\ref{C}) in terms of his state ${\bf r}$ as
\begin{equation}
|\cos(\,\eta_{{\bf a}{\bf e}_o})| \,\geqslant\, \cos\left(\frac{\pi}{2}r\right)
\;\;\text{and}\;\;\,
|\cos(\,\eta_{{\bf b}{\bf e}_o})| \,\geqslant\, \cos\left(\frac{\pi}{2}r\right)\!, \label{rcon}
\end{equation}
where our vector ${{\bf e}_o}$ is related to his vector ${\bf r}$ as ${\,{\bf e}_o={\bf r}/r}$.
We are thus treating the axis ${{\bf e}_o}$ and the angle ${\pi r}$ of the rotation of the spin as two independent random variables.

The equalities in the above inequalities correspond to the boundaries of the two circular caps on the spherical surface of radius ${r}$ within the SO(3) ball considered by Pearle. The intersection of the two circular caps is then
\begin{equation}
{\cal I}\!\left(\pi r,\,\eta_{{\bf a}{\bf b}}\right)=4r^2\!\!\int_{\frac{\eta_{{\bf a}{\bf b}}}{2}}^{\frac{\pi}{2}r}\!\!d\xi\sqrt{1-\left\{\!\frac{\cos\left(\frac{\pi}{2}r\right)}{\cos(\xi)}\!\right\}^2\,} \,\;\text{if}\;{\eta_{{\bf a}{\bf b}}\leqslant\pi r}, \label{area}
\end{equation}
and zero otherwise. This area is derived by Pearle in the Appendix A of his paper. It is, however, not the correct overlap area for our model. What has been overlooked in Pearle's derivation are the contributions to ${{\cal I}\!\left(\pi r,\,\eta_{{\bf a}{\bf b}}\right)}$ from the {\it relative} rotations of the state ${\,{\bf e}_o={\bf r}/r}$ along the directions ${\bf a}$ and ${{\bf b}}$. While the state ${{\bf e}_o}$ can be common to both ${\bf a}$ and ${\bf b}$, the corresponding rotations ${\pi r}$ cannot be the same in general about both ${\bf a}$ and ${\bf b}$. An example of the difference can be readily seen from the relations (\ref{f}) and (\ref{fp}), while heeding to the double covering in ${\mathrm{SU}(2)}$:
\begin{numcases}{\pi\Delta r=\!\!}
\!\!\!2\,\cos^{-1}\!\!\left[-1+\frac{2}{\sqrt{1+\,3\left(\frac{\eta_{{\bf a}{\bf b}}}{\pi}\right)}\,}\right] & \!\!\!\!\!\!\!\!\!\!\!${\text{if}\;\,0\leqslant \eta_{{\bf a}{\bf b}}\leqslant \frac{\pi}{2},}$ \notag \\
\\
\!\!2\,\cos^{-1}\!\!\left[-1+\frac{2}{\sqrt{4-\,3\left(\frac{\eta_{{\bf a}{\bf b}}}{\pi}\right)}\,}\right] & \!\!\!\!\!\!\!\!\!\!\!${\text{if}\;\,\frac{\pi}{2}\leqslant \eta_{{\bf a}{\bf b}} \leqslant \pi.}$ \notag
\end{numcases}
Evidently, ${\Delta r=0}$ when ${\eta_{{\bf a}{\bf b}}=0}$ or ${\pi}$, and maximum when ${\eta_{{\bf a}{\bf b}}\!=\!\frac{\pi}{2}}$. More generally, the effective radius of the spherical surface to which the circular caps belong must be ``phase-shifted" to ${r'=r\sqrt{{\rm h}(\eta_{{\bf a}{\bf b}})}}$ in our ${\mathrm{SU}(2)}$ model, where
\begin{equation}
{\rm h}(\eta_{{\bf a}{\bf b}})=\frac{3\pi}{8}\!\left\{\!\frac{\sin^2\!\left(\eta_{{\bf a}{\bf b}}\right)}{\pi\sin^2\!\left(\frac{1}{2}\eta_{{\bf a}{\bf b}}\right)+\eta_{{\bf a}{\bf b}}\cos\!\left(\eta_{{\bf a}{\bf b}}\right)-\sin\!\left(\eta_{{\bf a}{\bf b}}\right)}\!\right\} \label{h}
\end{equation}
is the inverse of the function derived in Pearle's Eq.~(23). The correct overlap area is then obtained by replacing ${r}$ by ${r'}$ in the differential area ${dA\!=\!r^2d\omega}$ in Eq.${\,}$(\ref{area}) so that
\begin{equation}
{\cal I}\!\left(\pi r,\,\eta_{{\bf a}{\bf b}}\right)\longrightarrow{\cal J}\!\left(\pi r,\,\eta_{{\bf a}{\bf b}}\right)={\rm h}(\eta_{{\bf a}{\bf b}})\,{\cal I}\!\left(\pi r,\,\eta_{{\bf a}{\bf b}}\right)\!. \label{del-I}
\end{equation}

Using the probability density (\ref{pdf}) and the overlap area (\ref{del-I}), we can now calculate various joint probabilities as
\begin{align}
P_{12}^{+-}(\eta_{{\bf a}{\bf b}})=P_{12}^{-+}(\eta_{{\bf a}{\bf b}})&=\!\!\int_{\frac{\eta_{{\bf a}{\bf b}}}{\pi}}^1 p(r)\;\frac{{\cal J}\!\left(\pi r,\,\eta_{{\bf a}{\bf b}}\right)}{4\pi r^2}\,dr \notag \\
&=\frac{1}{2}\cos^2\!\left(\frac{\eta_{{\bf a}{\bf b}}}{2}\right) \label{34}
\end{align}
and
\begin{align}
P_{12}^{++}(\eta_{{\bf a}{\bf b}})=P_{12}^{--}(\eta_{{\bf a}{\bf b}})&=\!\!\int_{1-\frac{\eta_{{\bf a}{\bf b}}}{\pi}}^1 p(r)\,\frac{{\cal J}\!\left(\pi r,\,\pi-\eta_{{\bf a}{\bf b}}\right)}{4\pi r^2}\,dr \notag \\
&=\frac{1}{2}\sin^2\!\left(\frac{\eta_{{\bf a}{\bf b}}}{2}\right)\!. \label{35}
\end{align}
These calculations of the joint probabilities are analogous to those by Pearle, except for using the area ${{\cal J}\!\left(\pi r,\,\eta_{{\bf a}{\bf b}}\right)}$. In particular, since ${{\rm h}(\eta_{{\bf a}{\bf b}})}$ expressed in (\ref{h}) is an inverse of the function derived in Pearle's Eq.~(23), our equations (\ref{34}) and (\ref{35}) follow at once from Pearle's equations (5) and (6), respectively, upon multiplying through with ${{\rm h}(\eta_{{\bf a}{\bf b}})}$.

Although the statistical effects of the constraints (\ref{rcon}) in our model turn out to be almost identical to those in Pearle's model, the characteristics of the two models are markedly different. In our model the vectors ${{\bf e}_o}$ and ${{\bf s}_o}$ ensure in tandem that there are no initial states for which
\begin{equation}
|\cos(\,\eta_{{\bf n}{\bf e}_o})| \,<\,\cos\left(\frac{\pi}{2} r\right)=
-1\,+\,\frac{2}{\sqrt{1+3\left(\frac{\eta_{{\bf z}{\bf s}_o}}{\kappa\pi}\right)}\,}. \label{Cleq}
\end{equation}
Consequently, the detectors of Alice and Bob can receive the spin states ${{\bf e}_o}$ only if the constraints (\ref{rcon}) are satisfied. In other words, unlike Pearle's model, our model is not concerned about data rejection or detection loophole. In particular, in our model the fraction ${{\rm g}(\eta_{{\bf a}{\bf b}})}$ of events in which both particles are detected is exactly equal to 1:
\begin{equation}
{\rm g}(\eta_{{\bf a}{\bf b}})\!=\frac{P_{12}^{+-}(\eta_{{\bf a}{\bf b}})}{\frac{1}{2}\cos^2\!\left(\frac{\eta_{{\bf a}{\bf b}}}{2}\right)}=\frac{P_{12}^{++}(\eta_{{\bf a}{\bf b}})}{\frac{1}{2}\sin^2\!\left(\frac{\eta_{{\bf a}{\bf b}}}{2}\right)}=1\;\forall\;\eta_{{\bf a}{\bf b}}\in[0,\,\pi]. \label{cl48}
\end{equation}
Clearly, a measurement event cannot occur if there does not exist a state which can bring about that event. Since the initial state of the system is specified by the pair ${({\bf e}_o,\,{\bf s}_o)}$ and not just by the vector ${{\bf e}_o}$, there are no states of the system for which ${|\cos(\,\eta_{{\bf n}{\bf e}_o})|\,<\,f(\eta_{{\bf z}{\bf s}_o})}$ for {\it any} vector ${\bf n}$. Thus a measurement event cannot occur for ${|\cos(\,\eta_{{\bf n}{\bf e}_o})|\,<\,f(\eta_{{\bf z}{\bf s}_o})}$, no matter what ${\bf n}$ is. As a result, there is a one-to-one correspondence between the initial state ${({\bf e}_o,\,{\bf s}_o)}$ selected from the set (\ref{lamset}) and the measurement events ${\mathscr A}$ and ${\mathscr B}$ specified by the Eqs.~(\ref{aamset}) and (\ref{bamset}). This means, in particular, that the ``fraction'' ${{\rm g}(\eta_{{\bf a}{\bf b}})}$ in our model is equal to 1 for all ${\eta_{{\bf a}{\bf b}}}$, dictating the vanishing of the probabilities
\begin{equation}
P_{12}^{00}(\eta_{{\bf a}{\bf b}})=1+{\rm g}(\eta_{{\bf a}{\bf b}})-2\,{\rm g}(0)=0,
\end{equation}
which follows from Pearle's Eq.${\,}$(9). Moreover, from his Eq.${\,}$(8) we also have ${P_{12}^{+0}(\eta_{{\bf a}{\bf b}})=\frac{1}{2}\left[\,{\rm g}(0)-{\rm g}(\eta_{{\bf a}{\bf b}})\right]}$, giving
\begin{equation}
P_{12}^{+0}(\eta_{{\bf a}{\bf b}})=P_{12}^{-0}(\eta_{{\bf a}{\bf b}})=P_{12}^{0+}(\eta_{{\bf a}{\bf b}})=P_{12}^{0-}(\eta_{{\bf a}{\bf b}})=0.
\end{equation}
Together with the probabilities for individual detections,  
\begin{equation}
P_1^{+}({\bf a})=P_1^{-}({\bf a})=P_2^{+}({\bf b})=P_2^{-}({\bf b})=\frac{1}{2}\,{\rm g}(0)=\frac{1}{2}\,,
\end{equation}
the correlation between ${\mathscr A}$ and ${\mathscr B}$ then works out to be
\begin{align}
\!\!\!\!{\cal E}({\bf a},\,{\bf b})\,&=\lim_{\,n\,\gg\,1}\left[\frac{1}{n}\sum_{i\,=\,1}^{n}\,
{\mathscr A}({\bf a};\,{\bf e}^i_o,\,{\bf s}_o^i)\;{\mathscr B}({\bf b};\,{\bf e}^i_o,\,{\bf s}_o^i)\right] \notag \\
&=\,\frac{P_{12}^{++}\,+\,P_{12}^{--}\,-\,P_{12}^{+-}\,-\,P_{12}^{-+}}{P_{12}^{++}\,+\,P_{12}^{--}\,+\,P_{12}^{+-}\,+\,P_{12}^{-+}}
\notag \\
&=\,-\,\cos\left(\eta_{{\bf a}{\bf b}}\right). \label{corsum}
\end{align}
Since all of the probabilities predicted by our local model in ${S^3}$ match exactly with the corresponding predictions of quantum mechanics, the violations of not only the CHSH inequality, but also Clauser-Horne inequality follow \cite{Clauser,disproof}.

\section{Event-by-event Numerical Simulations of the Strong Correlations}\label{simu}

We have verified the above results in several event-by-event numerical simulations \cite{Simulation-A,Simulation-B}, which provide further insights into the strength of the correlation for different values of ${\kappa}$. As we discussed above, the rotation angle ${\eta_{{\bf z}{\bf s}_o}\!}$ and the cumulative distribution function ${C(r)}$ are related by ${\kappa}$ as
\begin{equation}
\frac{\eta_{{\bf z}{\bf s}_o}}{\pi}=\kappa\,C(r),
\end{equation}
where ${|\kappa|\leqslant\infty}$ can be interpreted as a {\it strength constant}. It is easy to verify in the simulations \cite{Simulation-A,Simulation-B} that EPR-Bohm correlation results for ${\kappa=+1}$, whereas linear correlation results for ${\kappa=0}$. The unphysical, or PR box correlation can also be generated in the simulation by letting ${\kappa > +1}$. On the other hand, setting ${\kappa=-1}$ [which is equivalent to letting ${\eta_{{\bf z}{\bf s}_o}\!\rightarrow 2\pi-\eta_{{\bf z}{\bf s}_o}}$ in Eq.${\,}$(\ref{C})] leads back to the linear correlation \cite{Simulation-A,Simulation-B}. The crucial observation here is that the strong, or quantum correlations are manifested only for ${\kappa=+1}$. Consequently, they can be best understood as resulting from the geometrical and topological structures of the quaternionic ${S^3}$, as defined, for example, in Eq.~(\ref{onpara}).

This conclusion can be further substantiated by first reflecting on a non-quaternionic or vector representation of the 3-sphere to model rotations, and then returning back to the quaternionic representation to appreciate the difference. It is well known that tensors such as ordinary vectors are not capable of modelling rotations in the physical space, let alone modelling spinors in a singularity-free manner \cite{Christian,disproof}. However, in the present context we are not interested in modelling all possible rotations and their all possible compositions in the physical space. We are only interested in establishing the correct correlation between some very special limiting points of the 3-sphere, namely between its scalar points such as ${{\mathscr A}({\bf a},\,\lambda) = \pm 1}$ and ${{\mathscr B}({\bf b},\,\lambda) = \pm 1\;}$, with ${\lambda}$ being the ``hidden variable'' in the sense of Bell \cite{Bell-1964,Christian,disproof}. It turns out that in that case we can indeed model rotations (or more precisely, their spin values) by means of ordinary vectors and their inner products, but not with a single Riemannian metric \cite{Simulation-B}. A one-parameter family of {\it effective} metrics is required to model the relative spin values correctly. Given two vectors ${\bf u}$ and ${\bf v}$, their inner product ${g({\bf u},\,{\bf v},\,\eta)}$ is defined by the constraint ${\left|\,\cos({\bf u},\,{\bf v})\,\right| \geqslant f(\eta)\in[0,\,1]}$, with the two extreme cases, namely ${\left|\,\cos({\bf u},\,{\bf v})\,\right| \geqslant 0}$ and ${\left|\,\cos({\bf u},\,{\bf v})\,\right| \geqslant 1,}$ quantifying the weakest and the strongest topologies, respectively. Here the weakest topology dictated by ${\left|\,\cos({\bf u},\,{\bf v})\,\right| \geqslant 0}$ is the topology of ${{\mathrm{I\!R}^3}}$, where relatively few vectors ${\bf u}$ and ${\bf v}$ are orthogonal to each other. The strongest topology dictated by ${\left|\,\cos({\bf u},\,{\bf v})\right| \geqslant 1}$, on the other hand, is more interesting, since in that case nearly {\it all} of the vectors ${\bf u}$ and ${\bf v}$ are orthogonal to each other. All intermediate topologies are dictated by {\it the effective metric}
\begin{align}
&g({\bf u},\,{\bf v},\,\eta)=\begin{cases}{\bf u}\cdot{\bf v} & \text{if} \;\,\left|{\bf u}\cdot{\bf v}\right| \geqslant f(\eta) \\ 0 & \text{if}\;\,\left|{\bf u}\cdot{\bf v}\right| < f(\eta),\end{cases} \\
&\text{where}\,\;{\bf u}\cdot{\bf v} := \cos({\bf u},\,{\bf v})\,\;\text{and }\nonumber \\
&f(\eta)\,:=\,-1\,+\,\frac{2}{\sqrt{1+3\left(\frac{\eta}{\pi}\right)\,}\;}\,\;\;{\text with}\;\,\eta\in[0,\;\pi]\,. \label{ill}
\end{align}
Evidently, the orthogonality of the vectors ${\bf u}$ and ${\bf v}$ is defined here by the condition ${g({\bf u},\,{\bf v},\,\eta)=0}$, depending on the parameter ${\eta\,\in\,[0,\;\pi]}$. It is this one-parameter family of metrics that has been implemented in the simulations \cite{Simulation-A,Simulation-B}. The slight change in notation of the distribution function from that in Eq.${\,}$(\ref{C}) is purely for the coding convenience (cf. Fig.~\ref{fig3}).

\section{Derivation of the Strong Correlations using Geometric Algebra}\label{GA}

\begin{figure*}[t]
\hrule
\scalebox{1}{
\begin{pspicture}(4.5,-1.0)(5.0,5.9)
\psset{xunit=0.5mm,yunit=4cm}
\psaxes[axesstyle=frame,linewidth=0.01mm,tickstyle=full,ticksize=0pt,dx=90\psxunit,Dx=180,dy=1
\psyunit,Dy=+2,Oy=-1](0,0)(180,1.0)
\psline[linewidth=0.2mm,arrowinset=0.3,arrowsize=2pt 3,arrowlength=2]{->}(0,0.5)(190,0.5)
\psline[linewidth=0.2mm]{-}(45,0)(45,1)
\psline[linewidth=0.2mm]{-}(90,0)(90,1)
\psline[linewidth=0.2mm]{-}(135,0)(135,1)
\psline[linewidth=0.2mm,arrowinset=0.3,arrowsize=2pt 3,arrowlength=2]{->}(0,0)(0,1.2)
\psline[linewidth=0.35mm,linestyle=dashed,linecolor=gray]{-}(0,0)(90,1)
\psline[linewidth=0.35mm,linestyle=dashed,linecolor=gray]{-}(90,1)(180,0)
\put(2.1,-0.38){${90}$}
\put(6.5,-0.38){${270}$}
\put(-0.63,3.92){${+}$}
\put(-0.9,5.0){{\large ${{\cal E}^{\rm EPR}_{{\!}_{L.R.}}}$}${({\bf a},\,{\bf b})}$}
\put(-0.38,1.93){${0}$}
\put(9.65,1.95){\large ${\eta_{{\bf a}{\bf b}}}$}
\psplot[linewidth=0.35mm,linecolor=black]{0.0}{180}{x dup cos exch cos mul 1.0 mul neg 1 add}
\end{pspicture}}
\hrule
\caption{Plot of an event-by-event numerical simulation of the EPR-Bohm correlations predicted by our ${S^3}$ model \cite{{Simulation-A,Simulation-B,Wonnink,Diether-1,Diether-2}}. The x-axis depicts the angle in degrees between the measurement directions ${\bf a}$ and ${\bf b}$ chosen by Alice and Bob and the y-axis depicts the value of the correlation between their results.} 
\vspace{0.1cm}
\hrule
\label{fig3}
\end{figure*}

Returning to the singularity-free representation of ${S^3}$ specified in Eqs.${\,}$(\ref{onpara}) to (\ref{Q}), it is worth recalling that angular momenta are best described, not by ordinary polar vectors, but by pseudo-vectors, or bivectors, that change sign upon reflection \cite{Christian}. One only has to compare a spinning object, like a barber's pole, with its image in a mirror to appreciate this elementary fact. The mirror image of a polar vector representing the spinning object is not the polar vector that represents the mirror image of the spinning object. In fact it is the negative of the polar vector that does the job. Therefore the spin angular momenta considered previously are better represented by a set of unit bivectors using the powerful language of Geometric Algebra \cite{Clifford}. They can be expressed in terms of graded bivector bases with sub-algebra
\begin{equation}
L_{i}(\lambda)\,L_{j}(\lambda) \,=\,-\,\delta_{ij}\,-\,\sum_{k}\,\epsilon_{ijk}\,L_{k}(\lambda)\,, \label{wh-o8899}
\end{equation}
which span a tangent space at each point of ${S^3}$, with a choice of orientation ${\lambda=\pm\,1}$. Contracting this equation on both sides with the components ${a^{i}}$ and ${b^{j}}$ of arbitrary unit vectors ${\bf a}$ and ${\bf b}$ then gives the convenient bivector identity
\begin{equation}
{\bf L}({\bf a},\,\lambda)\,{\bf L}({\bf b},\,\lambda)\,=\,-\,{\bf a}\cdot{\bf b}\,-\,{\bf L}({\bf a}\times{\bf b},\,\lambda)\,, \label{50}
\end{equation}
where ${\mathbf{L}(\mathbf{a},\,\lambda):=a^iL_i(\lambda)}$ and ${\mathbf{L}(\mathbf{b},\,\lambda):=b^jL_j(\lambda)}$ are unit bivectors. Since ${\lambda}$ specifies the orientation of ${S^3}$ and not the handedness of a coordinate system [cf. Eq.~(\ref{un98})], the cross product ${{\bf a}\times{\bf b}}$ (which is of course universally defined by the right-hand rule) is not affected by it. The identity (\ref{50}) is simply a geometric product between the unit bivectors ${\mathbf{L}(\mathbf{a},\,\lambda)}$ and ${\mathbf{L}(\mathbf{b},\,\lambda)}$ representing the spin angular momenta considered previously:
\begin{align}
{\bf L}({\bf a},\,{\lambda})\,=\,\lambda\,I\,{\bf a}\,&=\,\lambda\,I\cdot{\bf a}\,\equiv\,\lambda({\bf e}_x{\bf e}_y{\bf e}_z)\cdot{\bf a}\notag \\
&=\,(\pm 1 \;\,\text{spin about the direction}\;\,{\bf a}) \label{12}
\end{align}
and\footnote{In Geometric Algebra \cite{Clifford} bivectors are viewed as {\it directed} numbers and characterized by only three abstract properties: (1) a magnitude, (2) a direction (specified by a vector orthogonal to it), and (3) a sense of rotation. The bivector ${{\bf L}({\bf a},\,\lambda)}$ thus specifies ${\pm 1}$ spin about the direction ${\bf a}$.}
\begin{align}
{\bf L}({\bf b},\,{\lambda})\,=\,\lambda\,I\,{\bf b}\,&=\,\lambda\,I\cdot{\bf b}\,\equiv\,\lambda({\bf e}_x{\bf e}_y{\bf e}_z)\cdot{\bf b} \notag \\
&=\,(\pm 1 \;\,\text{spin about the direction}\;\, {\bf b})\,, \label{13}
\end{align}
where the trivector ${I:=\,{\bf e}_x{\bf e}_y{\bf e}_z}$ with the property ${I^2=-1}$ represents the volume form on ${S^3}$ and ensures that ${{\bf L}^2({\bf n},\,\lambda)= -1}$. 

In the above representation of spins we have used algebra ${Cl_{3,0}}$ of the three-dimensional physical space \cite{Clifford}. However, as shown in Fig.~\ref{fig1}, in the EPR-Bohm type experiments the initial state ${\lambda}$ of the singlet system originates in the overlap of the backward light cones of Alice and Bob. The constituent spins emerging from the source are then detected at a spacelike distance from each other at a later time, as shown in Fig.~\ref{fig2}. Therefore, full relativistic considerations are essential in the analysis of local causality in the present context, as Bell has explained in Ref.~\cite{Bell-1990}. For that purpose, the appropriate algebra is ${Cl_{1,3}}$. It is also known as ``spacetime algebra", or STA \cite{Hestenes}. Within spacetime algebra ${Cl_{1,3}}$, we now represent the two constituent spins by using spacetime 4-vectors, as ${\lambda Is_1=\lambda Is^{\mu}_1\gamma_{\mu}}$ and ${\lambda Is_2=\lambda Is^{\mu}_2\gamma_{\mu}}$, where the set ${\{\gamma_0,\gamma_1,\gamma_2,\gamma_3\}}$ of ${\gamma}$-vectors forms an orthonormal basis, and its elements satisfy the same algebraic properties as Dirac matrices \cite{Hestenes}. The basis vectors ${\gamma_{\mu}}$ determine a unique pseudoscalar ${I=\gamma_0\gamma_1\gamma_2\gamma_3={\bf e}_x{\bf e}_y{\bf e}_z}$, with the properties ${I^2=-1}$ and ${\gamma_{\mu}I=-I\gamma_{\mu}}$. The two spins ${\lambda Is_1}$ and ${\lambda Is_2}$ are now spacetime covariant objects. They originate in the overlap of the backward light cones of Alice and Bob, along with the initial state ${\lambda}$, as shown in Fig.~\ref{fig2}.

We now wish to represent the measurements of these spins using the detectors ${Ia=Ia^{\mu}\gamma_{\mu}}$ and ${I\,b=I\,b^{\mu}\gamma_{\mu}}$, which we also represent using 4-vectors in ${Cl_{1,3}}$, situated at a spacelike distance from each other, within the hypersurface ${S^3}$. Such a spacelike hypersurface can be specified in ${Cl_{1,3}}$ using the timelike vector ${\gamma_0}$, representing our two observers:
\begin{align}
Ia\,\gamma_0 &= Ia^{\mu}\gamma_{\mu}\gamma_0 = I a^0{\gamma}_0{\gamma}_0 + Ia^i\gamma_i\gamma_0 \notag \\
&= a^0 I + Ia^i{\bf e}_i = a^0 I + I\cdot{\bf a} = a^0 I + {\bf D}({\bf a})
\end{align}
and
\begin{align}
\;I\,b\,\gamma_0 &= I\,b^{\mu}\gamma_{\mu}\gamma_0 = I\,b^0{\gamma}_0{\gamma}_0 + I\,b^i\gamma_i\gamma_0 \notag \\
&= b^0 I + I\,b^i{\bf e}_i = b^0 I + I\cdot{\bf b} = b^0 I + {\bf D}({\bf b})\,, 
\end{align}
because ${{\gamma}_0^2=1}$ and ${\gamma_i\gamma_0 = {\bf e}_i}$ constitute the spatial basis vectors in the algebra ${Cl_{3,0}}$ \cite{Hestenes}. Such an observer-dependent projection of a 4-vector in ${Cl_{1,3}}$ onto a (1+3)-dimensional space, resulting in a timelike scalar and a spacelike vector in ${Cl_{3,0}}$, is called a ``spacetime split" in STA. Here ${I\cdot{\bf a}}$ $=$ ${{\bf D}({\bf a})}$ and ${I\cdot{\bf b}}$ $=$ ${{\bf D}({\bf b})}$ are spacelike bivectors in ${Cl_{3,0}}$, analogous to those in Eqs.~(\ref{12}) and (\ref{13}). We will use ${{\bf D}({\bf a})}$ and ${{\bf D}({\bf b})}$ to represent the two detectors of Alice and Bob.

Now, since the spins are measured by the above detectors, we must project them as well onto the hypersurface ${S^3}$:
\begin{align}
\lambda Is_1\gamma_0 &= \lambda Is_1^{\mu}\gamma_{\mu}\gamma_0 = \lambda Is_1^0{\gamma}_0{\gamma}_0 + \lambda Is_1^i\gamma_i\gamma_0 \notag \\
&= \lambda Is_1^0 + \lambda Is_1^i{\bf e}_i = \lambda Is_1^0 + \lambda I\cdot{\bf s}_1 \notag \\
&= \lambda Is_1^0 + {\bf L}({\bf s}_1,\,\lambda)
\end{align}
and
\begin{align}
\lambda Is_2\gamma_0 &= \lambda Is_2^{\mu} \gamma_{\mu}\gamma_0 = \lambda Is_2^0{\gamma}_0{\gamma}_0 + \lambda Is_2^i\gamma_i\gamma_0 \notag \\
&= \lambda Is_2^0 + \lambda Is_2^i{\bf e}_i = \lambda Is_2^0 + \lambda I\cdot{\bf s}_2 \notag \\
&= \lambda Is_2^0 + {\bf L}({\bf s}_2,\,\lambda)\,.
\end{align}
Next, we set up the time coordinate from ${-t}$ to ${0}$ in Figs.~\ref{fig1} and \ref{fig2}, where ${-t}$ is the instant at which the singlet state is produced from the source in the overlap of the backward light cones of Alice and Bob, and ${0}$ is the instant at which the constituent spins ${\lambda Is_1^0 + {\bf L}({\bf s}_1,\,\lambda)}$ and ${\lambda Is_2^0 + {\bf L}({\bf s}_2,\,\lambda)}$ are measured {\it simultaneously} by the detectors ${a^0 I + {\bf D}({\bf a})}$ and ${b^0 I + {\bf D}({\bf b})}$, respectively. Consequently, since ${a^0}$, ${b^0}$, ${s_1^0}$, and ${s_2^0}$ can be identified as the respective time coordinates, in the coordinates in which the spins are measured simultaneously by Alice and Bob we must set ${a^0=b^0=s_1^0=s_2^0=0}$, reducing the spin and detector 4-vectors within ${Cl_{1,3}}$ to the bivectors ${{\bf L}({\bf s}_1,\,\lambda)}$, ${{\bf L}({\bf s}_2,\,\lambda)}$, ${{\bf D}({\bf a})}$, and ${{\bf D}({\bf b})}$ within ${Cl_{3,0}}$. The two relativistic spins ${\lambda Is_1}$ and ${\lambda Is_2}$ that originate in the overlap of the backward light cones of Alice and Bob at time ${-t}$ are thus detected as ${{\bf L}({\bf s}_1,\,\lambda)}$ and ${{\bf L}({\bf s}_2,\,\lambda)}$ at time ${0}$ by the detectors ${{\bf D}({\bf a})}$ and ${{\bf D}({\bf b})}$, respectively. 

We are now in a position to derive the singlet correlation once again in a succinct and elegant manner. To this end, consider two measurement functions similar to (\ref{nomap-1-2jim}) and (\ref{nomap-1-2job}), but now of the form
\begin{equation}
\pm\,1\,=\,{\mathscr A}({\bf a},\,\lambda^k)\!:
{\mathrm{I\!R}}^3\!\times\left\{\,\lambda^k\right\}\longrightarrow S^3\hookrightarrow{\mathrm{I\!R}}^4
\end{equation}
and
\begin{equation}
\pm\,1\,=\,{\mathscr B}({\bf b},\,\lambda^k)\!:
{\mathrm{I\!R}}^3\!\times\left\{\,\lambda^k\right\}\longrightarrow S^3\hookrightarrow{\mathrm{I\!R}}^4,\end{equation}
where ${\lambda^k=\pm1}$ for each run ${k}$ of the experiment considered in Fig.~\ref{fig1}. More explicitly, let the spin bivectors ${\mp\,{\bf L}({\bf s},\,\lambda^k)}$ emerging from a common source be detected by two space-like separated detector bivectors ${{\bf D}({\bf a})}$ and ${{\bf D}({\bf b})}$, giving
\begin{align}
S^3\ni\,{\mathscr A}({\bf a},\,{\lambda^k})\,:&=\,\lim_{{\bf s}_1\,\rightarrow\,{\bf a}}\left\{-\,{\bf D}({\bf a})\,{\bf L}({\bf s}_1,\,\lambda^k)\right\} \notag \\
&=\,                  
\begin{cases}
+\,1\;\;\;\;\;{\rm if} &\lambda^k\,=\,+\,1 \\
-\,1\;\;\;\;\;{\rm if} &\lambda^k\,=\,-\,1
\end{cases} \Bigg\} \notag \\
&\;\;\;\;\;\;\;\text{with}\;\, \Bigl\langle\,{\mathscr A}({\bf a},\,\lambda^k)\,\Bigr\rangle\,=\,0 \label{53}
\end{align}
and
\begin{align}
S^3\ni\,{\mathscr B}({\bf b},\,{\lambda^k})\,:&=\,\lim_{{\bf s}_2\,\rightarrow\,{\bf b}}\left\{+\,{\bf L}({\bf s}_2,\,\lambda^k)\,{\bf D}({\bf b})\right\} \notag \\
&=\,                     
\begin{cases}
-\,1\;\;\;\;\;{\rm if} &\lambda^k\,=\,+\,1 \\
+\,1\;\;\;\;\;{\rm if} &\lambda^k\,=\,-\,1
\end{cases} \Bigg\} \notag \\
&\;\;\;\;\;\;\;\text{with}\;\,\Bigl\langle\,{\mathscr B}({\bf b},\,\lambda^k)\,\Bigr\rangle\,=\,0\,, \label{54}
\end{align}
where we assume the orientation ${\lambda}$ of ${S^3}$ to be a random variable with 50/50 chance of being ${+1}$ or ${-\,1}$ at the moment of the pair-creation, making the spinning bivector ${{\bf L}({\bf n},\,\lambda^k)}$ a random variable {\it relative} to the detector bivector ${{\bf D}({\bf n})}$:
\begin{equation}
{\bf L}({\bf n},\,\lambda^k)\,=\,\lambda^k\,{\bf D}({\bf n})\,\,\Longleftrightarrow\,\,{\bf D}({\bf n})\,=\,\lambda^k\,{\bf L}({\bf n},\,\lambda^k)\,. \label{55}
\end{equation}
It is important to recall here that an orientation of a manifold is a {\it relative} concept \cite{RSOS}. Within Geometric Algebra ${Cl_{3,0}}$ the choice of the sign of the unit pseudoscalar amounts to choosing an orientation of the physical space \cite{Clifford,Dorst}.

It is evident from the measurement functions ${{\mathscr A}({\bf a},\,{\lambda^k})}$ and ${{\mathscr B}({\bf b},\,{\lambda^k})}$ defined above that their values are limiting scalar points, ${\pm1}$, of some quaternions that constitute ${S^3}$. Consequently, they respect the topology of ${S^3}$ rather than that of ${{\mathrm{I\!R}^3}}$. Physically, the geometry and topology of ${S^3}$ arise from the rotations of the two spin bivectors, ${-{\bf L}({\bf s}_1,\,\lambda^k)}$ and ${+{\bf L}({\bf s}_2,\,\lambda^k)}$, {\it relative} to the detector bivectors ${{\bf D}({\bf a})}$ and ${{\bf D}({\bf b})}$, respectively. This is evident from the definitions of the measurement functions, which involve geometric products of the form ${-{\bf D}({\bf a}){\bf L}({\bf s}_1,\,\lambda^k)}$, which are non-pure quaternions, and therefore elements of the set ${S^3}$. In other words, the interactions between the spin bivectors and the detector bivectors are represented in the model by non-pure quaternions of the form ${-{\bf D}({\bf a}){\bf L}({\bf s}_1,\,\lambda^k)}$, which --- as constituents of the set ${S^3}$ --- necessarily capture the geometry and topology of ${S^3}$.

Despite being algebraically different expressions, the detection processes encoded by  Eqs.~(\ref{53}) and (\ref{54}) are effectively the same as those defined by Bell in his local model~\cite{Bell-1964,Peres} within ${{\mathrm{I\!R}}^3}$, namely ${{\mathscr A}({\bf a},\,{\lambda^k})\cong {\mathrm{sign}}(+\,{\bf s}^k_1\cdot{\bf a})=\pm1}$ and ${{\mathscr B}({\bf b},\,{\lambda^k})\cong {\mathrm{sign}}(-\,{\bf s}^k_2\cdot{\bf b})=\pm1}$. They pick out the normalized components of the two spins about the vectors ${\bf a}$ and ${\bf b}$, representing two scalar points of the 3-sphere. To see this explicitly, we can expand the RHS of Eq.${\,}$(\ref{53}) as follows:  
\begin{align}
\lim_{{\bf s}_1\,\rightarrow\,{\bf a}}&\left\{-\,{\bf D}({\bf a})\,{\bf L}({\bf s}_1,\,\lambda^k)\right\} \notag \\
&\,=\,\lim_{{\bf s}_1\,\rightarrow\,{\bf a}}\left\{-\,\lambda^k\,{\bf L}({\bf a},\,\lambda^k)\,{\bf L}({\bf s}_1,\,\lambda^k)\right\} \\
&\,=\,\lim_{{\bf s}_1\,\rightarrow\,{\bf a}}\left[\,-\lambda^k\left\{-\,{\bf a}\cdot{\bf s}_1\,-\,{\bf L}({\bf a}\times{\bf s}_1,\,\lambda^k)\right\}\right] \\
&\,=\,\lim_{{\bf s}_1\,\rightarrow\,{\bf a}}\left\{\,+\,{\bf a}\cdot(\lambda^k\,{\bf s}_1)\,+\,I\cdot({\bf a}\times{\bf s}_1)\right\} \\
&\,\cong\,{\mathrm{sign}}(+\,{\bf s}^k_1\cdot{\bf a})\,,\;\;\text{with}\;\;{\bf s}^k_1\equiv\lambda^k\,{\bf s}_1, \label{600}
\end{align}
where Eqs.${\,}$(\ref{55}), (\ref{50}), and (\ref{12}) are used. Likewise we can expand the RHS of Eq.${\,}$(\ref{54}) using Eqs.${\,}$(\ref{55}), (\ref{50}), and (\ref{13}):
\begin{align}
\lim_{{\bf s}_2\,\rightarrow\,{\bf b}}&\left\{+\,{\bf L}({\bf s}_2,\,\lambda^k)\,{\bf D}({\bf b})\right\} \notag \\
&\,=\,\lim_{{\bf s}_2\,\rightarrow\,{\bf b}}\left\{+\,{\bf L}({\bf s}_2,\,\lambda^k)\,\lambda^k\,{\bf L}({\bf b},\,\lambda^k)\right\} \\
&\,=\,\lim_{{\bf s}_2\,\rightarrow\,{\bf b}}\left[\,+\lambda^k\left\{-\,{\bf s}_2\cdot{\bf b}\,-\,{\bf L}({\bf s}_2\times{\bf b},\,\lambda^k)\right\}\right] \\
&\,=\,\lim_{{\bf s}_2\,\rightarrow\,{\bf b}}\left\{\,-\,(\lambda^k\,{\bf s}_2)\cdot{\bf b}\,-\,I\cdot({\bf s}_2\times{\bf b})\right\} \\
&\,\cong\,{\mathrm{sign}}(-\,{\bf s}^k_2\cdot{\bf b})\,,\;\;\text{with}\;\;{\bf s}^k_2\equiv\lambda^k\,{\bf s}_2.
\end{align}
Moreover, as demanded by the conservation of angular momentum, we require the total spin to respect the condition
\begin{align}
-\,{\bf L}({\bf s}_1,\,\lambda^k)\,&+\,{\bf L}({\bf s}_2,\,\lambda^k)\,=\,0 \notag \\
&\Longleftrightarrow\;\; {\bf L}({\bf s}_1,\,\lambda^k)\,=\,{\bf L}({\bf s}_2,\,\lambda^k) \notag \\
&\Longleftrightarrow\;\; {\bf s}_1=\,{\bf s}_2\,\equiv\,{\bf s}\;
\;\;\,[\text{cf. Fig.~\ref{fig1}}]. \label{56}
\end{align}
Evidently, in the light of the product rule (\ref{50}) for the unit bivectors, the above condition is equivalent to the condition
\begin{equation}
{\bf L}({\bf s}_1,\,\lambda^k)\,{\bf L}({\bf s}_2,\,\lambda^k)=\left\{\,{\bf L}({\bf s},\,\lambda^k)\right\}^2=\,{\bf L}^2({\bf s},\,\lambda^k)=-1\,. \label{566}
\end{equation}
Note, however, that the limits ${\mathbf{s}_1\to \mathbf{a}}$ and ${\mathbf{s}_2\to \mathbf{b}}$ appearing in the definitions of the two measurement functions (\ref{53}) and (\ref{54}) are parts of the {\it independent} detection processes. These processes are {\it not} subject to the conservation law dictated by Eq.~(\ref{56}) or (\ref{566}), which remains valid only for the free evolution of the constituent spins. In fact, the detection processes describe purely local interactions of the spin bivectors with the detector bivectors, occurring at spacelike separated observation stations of Alice and Bob (see answer 9 in Appendix B for further explanation). Consequently, the expectation value of the simultaneous measurement outcomes ${{\mathscr A}({\bf a},\,{\lambda^k})=\pm1}$ and ${{\mathscr B}({\bf b},\,{\lambda^k})=\pm1}$ in ${S^3}$ works out as follows:
\begin{widetext}
\begin{align}
{\cal E}({\bf a},\,{\bf b})\,&=\lim_{\,n\,\rightarrow\,\infty}\left[\frac{1}{n}\sum_{k\,=\,1}^{n}\,{\mathscr A}({\bf a},\,{\lambda}^k)\;{\mathscr B}({\bf b},\,{\lambda}^k)\right] \,\cong\lim_{\,n\,\rightarrow\,\infty}\left[\frac{1}{n}\sum_{k\,=\,1}^{n}\,
{\mathrm{sign}}(+\,{\bf s}^k_1\cdot{\bf a})\;{\mathrm{sign}}(-\,{\bf s}^k_2\cdot{\bf b})\right]\label{57}\\
&=\lim_{\,n\,\rightarrow\,\infty}\left[\frac{1}{n}\sum_{k\,=\,1}^{n}\,\bigg[\lim_{{\bf s}_1\,\rightarrow\,{\bf a}}\left\{\,-\,{\bf D}({\bf a})\,{\bf L}({\bf s}_1,\,\lambda^k)\right\}\bigg]\left[\lim_{{\bf s}_2\,\rightarrow\,{\bf b}}\left\{\,+\,{\bf L}({\bf s}_2,\,\lambda^k)\,{\bf D}({\bf b})\right\}\,\right]\right] \label{58}\\
&=\lim_{\,n\,\rightarrow\,\infty}\left[\frac{1}{n}\sum_{k\,=\,1}^{n}\,\lim_{\substack{{\bf s}_1\,\rightarrow\,{\bf a} \\ {\bf s}_2\,\rightarrow\,{\bf b}}}\left\{\,-\,{\bf D}({\bf a})\,\right\}\,\left\{\,{\bf L}({\bf s}_1,\,\lambda^k)\,\,{\bf L}({\bf s}_2,\,\lambda^k)\,\right\}\,\left\{\,+\,{\bf D}({\bf b})\,\right\}\right] \label{59} \\
&=\lim_{\,n\,\rightarrow\,\infty}\left[\frac{1}{n}\sum_{k\,=\,1}^{n}\,\lim_{\substack{{\bf s}_1\,\rightarrow\,{\bf a} \\ {\bf s}_2\,\rightarrow\,{\bf b}}}\left\{-\,\lambda^k\,{\bf L}({\bf a},\,\lambda^k)\right\}\,\left\{\,-1\,\right\}\,\left\{+\,\lambda^k\,{\bf L}({\bf b},\,\lambda^k)\right\}\right] \label{60}\\
&=\lim_{\,n\,\rightarrow\,\infty}\left[\frac{1}{n}\sum_{k\,=\,1}^{n}\,\lim_{\substack{{\bf s}_1\,\rightarrow\,{\bf a} \\ {\bf s}_2\,\rightarrow\,{\bf b}}}\left\{\,+\,\left(\lambda^k\right)^2\,{\bf L}({\bf a},\,\lambda^k)\,\,{\bf L}({\bf b},\,\lambda^k)\right\}\right] \label{61} \\
&=\lim_{\,n\,\rightarrow\,\infty}\left[\frac{1}{n}\sum_{k\,=\,1}^{n}\,{\bf L}({\bf a},\,\lambda^k)\,{\bf L}({\bf b},\,\lambda^k)\,\right] \label{62}\\
&=\,-\,{\bf a}\cdot{\bf b}\,-\!\lim_{\,n\,\rightarrow\,\infty}\left[\frac{1}{n}\sum_{k\,=\,1}^{n}\,{\bf L}({\bf a}\times{\bf b},\,\lambda^k)\,\right] \label{63}\\
&=\,-\,{\bf a}\cdot{\bf b}\,-\!\lim_{\,n\,\rightarrow\,\infty}\left[\frac{1}{n}\sum_{k\,=\,1}^{n}\,\lambda^k\,\right]{\bf D}({\bf a}\times{\bf b}) \label{64}\\
&=\,-\,{\bf a}\cdot{\bf b}\,+\,0\,. \label{65}
\end{align}
\end{widetext}
Here Eq.${\,}$(\ref{58}) follows from Eq.${\,}$(\ref{57}) by substituting the functions ${{\mathscr A}({\bf a},\,{\lambda^k})}$ and ${{\mathscr B}({\bf b},\,{\lambda^k})}$ from the definitions (\ref{53}) and (\ref{54}); Eq.${\,}$(\ref{59}) follows from Eq.${\,}$(\ref{58}) by using the ``product of limits equal to limits of product'' rule [which can be verified by recognizing that the same quaternion ${-\,{\bf D}({\bf a})\,{\bf L}({\bf a},\,\lambda^k)\,{\bf L}({\bf b},\,\lambda^k)\,{\bf D}({\bf b})}$ results from the limits in Eqs.${\,}$(\ref{58}) and (\ref{59})]; Eq.${\,}$(\ref{60}) follows from Eq.${\,}$(\ref{59}) by (i) using the relation (\ref{55}) [thus setting all bivectors in the spin bases], (ii) the associativity of the geometric product, and (iii) the conservation of spin angular momentum specified in Eq.${\,}$(\ref{566}); Eq.${\,}$(\ref{61}) follows from Eq.${\,}$(\ref{60}) by recalling that scalars such as ${\lambda^k}$ commute with the bivectors; Eq.${\,}$(\ref{62}) follows from Eq.${\,}$(\ref{61}) by using ${\lambda^2 = +1}$, and by removing the superfluous limit operations; Eq.${\,}$(\ref{63}) follows from Eq.${\,}$(\ref{62}) by using the geometric product or identity (\ref{50}), together with the fact that there is no third spin about the orthogonal direction ${{\bf a}\times{\bf b}}$ once the two spins are already detected along the directions ${\bf a}$ and ${\bf b}$; Eq.${\,}$(\ref{64}) follows from Eq.${\,}$(\ref{63}) by using the relations (\ref{55}) and summing over the counterfactual detections of the ``third'' spins about ${{\bf a}\times{\bf b}}$; and Eq.${\,}$(\ref{65}) follows from Eq.${\,}$(\ref{64}) because the scalar coefficient of the bivector ${{\bf D}({\bf a}\times{\bf b})}$ vanishes in ${n\rightarrow\infty}$ limit, since ${\lambda^k}$ is a fair coin.

Note that, apart from the initial state ${\lambda^k}$, the only other assumption used in this derivation is that of the conservation of spin angular momentum (\ref{566}). These two assumptions are necessary and sufficient to dictate the singlet correlations:
\begin{equation}
{\cal E}({\bf a},\,{\bf b})\,=\lim_{\,n\,\rightarrow\,\infty}\left[\frac{1}{n}\sum_{k\,=\,1}^{n}\,{\mathscr A}({\bf a},\,{\lambda}^k)\;{\mathscr B}({\bf b},\,{\lambda}^k)\right]\!
=-\,{\bf a}\cdot{\bf b}. \label{abcos}
\end{equation}
This demonstrates that EPR-Bohm correlations are correlations among the scalar points of a quaternionic 3-sphere\footnote{\label{Hardy}The singlet correlations reproduced here can be reproduced also within a more general 7-sphere framework without any reference to the 3-sphere model presented here, as we have shown elsewhere \cite{RSOS}. And even though the highly non-trivial Hardy-type correlations can also be reproduced within the quaternionic 3-sphere model presented here (as we have demonstrated in Chapter 6 of Ref.~\cite{disproof}), the quaternionic 3-sphere model is rather restrictive. It can accommodate the singlet correlations and Hardy-type correlations, but cannot reproduce more intricate correlations, such as, for example, those predicted by the rotationally non-invariant GHZ states \cite{disproof}. On the other hand, the 7-sphere framework of Ref.~\cite{RSOS} is more general and comprehensive. This is because ${S^7}$ is made of ${S^4}$ worth of 3-spheres, with a highly non-trivial twist in the bundle. In other words, in the language of Hopf fibration, ${S^7}$ is fibrated by ${S^3}$ over the base manifold ${S^4}$. Thus each of the many fibers of ${S^7}$ that make it up is itself an ${S^3}$. It is therefore not surprising that ${S^7}$ framework is more complete and is able to reproduce quantum correlations more comprehensively. Moreover, as shown in Ref.~\cite{RSOS}, the algebraic and geometrical properties of the physical space are captured more completely by the octonion-like representation space ${S^7}$ constructed in Ref.~\cite{RSOS}, rather than by the 3-dimensional closed and compact physical space ${S^3}$ itself.}.

It is also instructive to evaluate the sum in Eq.${\,}$(\ref{62}) somewhat differently to bring out how the orientation ${\lambda^k}$ plays an important role in the derivation of the above correlation. Instead of assuming ${\lambda^k=\pm1}$ to be an orientation of ${S^3}$, we may view it as specifying the ordering relation between the spin bivectors ${{\bf L}({\bf a},\,\lambda)}$ and ${{\bf L}({\bf b},\,\lambda)}$ and the detector bivectors ${{\bf D}({\bf a}})$ and ${{\bf D}({\bf b})}$ with 50/50 chance of occurring, and only subsequently identify it with an orientation of ${S^3}$:
\begin{equation}
{\bf L}({\bf a},\,{\lambda}=+1)\;{\bf L}({\bf b},\,{\lambda}=+1)\,=\,{\bf D}({\bf a})\;{\bf D}({\bf b}) \label{pair1}
\end{equation}
or
\begin{equation}
\text{\;}{\bf L}({\bf a},\,{\lambda}=-1)\;{\bf L}({\bf b},\,{\lambda}=-1)\,=\,{\bf D}({\bf b})\;{\bf D}({\bf a}). \label{pair2}
\end{equation}
Since the spins emerging from the source are oblivious to the detectors located at remote stations, we may represent spins with a trivector ${J}$ and detectors with a trivector ${I}$, respectively, without assuming any relation between them:
\begin{equation}
{\bf L}({\bf n},\,{\lambda})\,=\,J\cdot{\bf n} \label{JB}
\end{equation}
and
\begin{equation}
\text{\;}\;{\bf D}({\bf n})\,=\,I\cdot{\bf n}\,, \label{IB}
\end{equation}
for any given dual vector ${\bf n}$. We can now easily find the relationship between ${J}$ and ${I}$ using the identities (\ref{wh-o8899}) and 
\begin{equation}
{\bf D}({\bf a})\,{\bf D}({\bf b})\,=\,-\,{\bf a}\cdot{\bf b}\,-\,{\bf D}({\bf a}\times{\bf b})\,. \label{bbA5000}
\end{equation}
Substituting the right-hand sides of these identities into the ordering relations (\ref{pair1}) and (\ref{pair2}) reduces the relations to
\begin{equation}
-\,{\bf a}\cdot{\bf b}\,-\,{\bf L}({\bf a}\times{\bf b},\,\lambda=+1)\,=\,-\,{\bf a}\cdot{\bf b}\,-\,{\bf D}({\bf a}\times{\bf b})
\end{equation}
or
\begin{align}
-\,{\bf a}\cdot{\bf b}\,-\,{\bf L}({\bf a}\times{\bf b},\,\lambda=-1)\,
&=\,-\,{\bf b}\cdot{\bf a}\,-\,{\bf D}({\bf b}\times{\bf a}) \notag \\ 
&=\,-\,{\bf a}\cdot{\bf b}\,+\,{\bf D}({\bf a}\times{\bf b})\,,
\end{align}
which, after canceling the scalar factor ${-\,{\bf a}\cdot{\bf b}}$ and using ${\lambda=\pm1}$ and the definitions (\ref{JB}) and (\ref{IB}), further reduces${\;}$to
\begin{align}
{\bf L}({\bf a}\times{\bf b},\,\lambda)\,
&=\,\lambda\,{\bf D}({\bf a}\times{\bf b}) \label{sit84}\\
J\cdot({\bf a}\times{\bf b})\,
&=\,\lambda\,I\cdot({\bf a}\times{\bf b}) \\
J\,
&=\,\lambda\,I\,. \label{un98}
\end{align}
We have thus proved that the ordering relations (\ref{pair1}) and (\ref{pair2}) between the spin bivectors ${{\bf L}({\bf a},\,\lambda)}$ and ${{\bf L}({\bf b},\,\lambda)}$ and the detector bivectors ${{\bf D}({\bf a}})$ and ${{\bf D}({\bf b})}$ are equivalent to our hypothesis that the orientation of the 3-sphere is a fair coin. Using the relations (\ref{50}) and (\ref{sit84}), together with the ordering relations (\ref{pair1}) and (\ref{pair2}), the sum (\ref{62}) can now be evaluated directly by recognizing that the spins in the right and left oriented ${S^3}$ satisfy the following geometrical relations \cite{Christian,disproof}:
\begin{align}
{\bf L}({\bf a},\,{\lambda}^k=+1)\;{\bf L}({\bf b},\,{\lambda}^k=+1)\,&=\,-\,{\bf a}\cdot{\bf b}\,-\,{\bf D}({\bf a}\times{\bf b}) \nonumber \\
&=\,{\bf D}({\bf a})\;{\bf D}({\bf b}) \notag \\
&=\,(\,+\,I\cdot{\bf a})(\,+\,I\cdot{\bf b}) \label{87}
\end{align}
and
\begin{align}
{\bf L}({\bf a},\,{\lambda}^k=-1)\;{\bf L}({\bf b},\,{\lambda}^k=-1)\,&=\,-\,{\bf a}\cdot{\bf b}\,+\,{\bf D}({\bf a}\times{\bf b}) \notag \\
&=\,-\,{\bf b}\cdot{\bf a}\,-\,{\bf D}({\bf b}\times{\bf a}) \nonumber \\
&=\,{\bf D}({\bf b})\;{\bf D}({\bf a}) \notag \\
&=\,(\,+\,I\cdot{\bf b})(\,+\,I\cdot{\bf a}). \label{88}
\end{align}
In other words, when the initial state ${\lambda^k}$ happens to be equal to ${+1}$, ${{\bf L}({\bf a},\,{\lambda}^k)\;{\bf L}({\bf b},\,{\lambda}^k)=(\,+\,I\cdot{\bf a})(\,+\,I\cdot{\bf b})}$, and when the initial state ${\lambda^k}$ happens to be equal to ${-1}$, ${\,{\bf L}({\bf a},\,{\lambda}^k)\;{\bf L}({\bf b},\,{\lambda}^k)=(\,+\,I\cdot{\bf b})(\,+\,I\cdot{\bf a})}$. Consequently, the expectation value (\ref{57}) reduces at once to
\begin{align}
{\cal E}({\bf a},\,{\bf b})\,&=\,\frac{1}{2}(\,+\,I\cdot{\bf a})(\,+\,I\cdot{\bf b})\,+\,\frac{1}{2}(\,+\,I\cdot{\bf b})(\,+\,I\cdot{\bf a}) \notag \\ 
&=\,-\,\frac{1}{2}\left\{{\bf a}{\bf b}\,+\,{\bf b}{\bf a}\right\}\,=\,-\,{\bf a}\cdot{\bf b}\,+\,0\,,\label{stand-nossss}
\end{align}
because the orientation ${\lambda^k}$ of ${S^3}$ is a fair coin. Here the last equality follows from the definition of the inner product. Given this result, it is not difficult to derive the corresponding upper bound on the expectation values within ${S^3}$, as we have demonstrated in Section \ref{apenB}:
\begin{equation}
\left|\,{\cal E}({\bf a},\,{\bf b})\,+\,{\cal E}({\bf a},\,{\bf b'})\,+\,{\cal E}({\bf a'},\,{\bf b})\,-\,{\cal E}({\bf a'},\,{\bf b'})\,\right|\,\leqslant\,2\sqrt{2}\,. \label{My-CHSH}
\end{equation}
We have verified both of these results in several numerical simulations \cite{Simulation-A,Simulation-B,Wonnink,Diether-1,Diether-2,Chantal}. The simulations are instructive on their own right, and some of them can be used for testing the effects of topology changes when the parameter ${\eta\in[0,\;\pi]}$ is varied.

\section{The Raison D'\^etre of Strong Correlations}\label{raison}

\begin{figure*}[t]
\hrule
\scalebox{0.8}{
\begin{pspicture}(0.3,-3.9)(4.5,3.1)

\pscircle[linewidth=0.3mm,linestyle=dashed](-1.8,-0.45){2.6}

\psellipse[linewidth=0.3mm](-0.8,-0.45)(0.7,1.4)

\psellipse[linewidth=0.3mm,border=3pt](-2.4,-0.45)(1.4,0.4)

\pscurve[linewidth=0.3mm,border=3pt](-1.485,-0.35)(-1.48,-0.25)(-1.45,0.0)

\pscircle[linewidth=0.3mm](7.0,-0.45){1.7}

\psellipse[linewidth=0.2mm,linestyle=dashed](7.0,-0.45)(1.68,0.4)

\put(-4.4,1.27){{\Large ${S^3}$}}

\put(-2.0,1.2){{\Large ${h^{-1}(q)}$}}

\put(-3.7,-1.4){{\Large ${h^{-1}(p)}$}}

\put(7.43,0.67){{\Large ${q}$}}

\psdot*(7.2,0.79)

\put(6.3,0.43){{\Large ${p}$}}

\psdot*(6.1,0.43)

\put(8.5,-1.8){{\Large ${S^2}$}}

\put(6.1,-2.7){\large base space}

\put(1.9,0.7){\Large ${h:S^3\rightarrow S^2}$}

\pscurve[linewidth=0.3mm,arrowinset=0.3,arrowsize=3pt 3,arrowlength=2]{->}(1.2,0.25)(2.47,0.45)(3.74,0.45)(4.9,0.25)

\put(1.73,-0.45){\large Hopf fibration}

\pscurve[linewidth=0.3mm,arrowinset=0.3,arrowsize=3pt 3,arrowlength=2]{->}(4.9,-0.95)(3.74,-1.15)(2.47,-1.15)(1.2,-0.95)

\put(1.75,-1.8){\Large ${h^{-1}:S^2\rightarrow S^3}$}

\end{pspicture}}
\hrule
\caption{The tangled web of linked Hopf circles depicting the geometrical and topological non-trivialities of the 3-sphere. Locally (in the topological sense) ${S^3}$ can be written as the product ${S^2\times S^1}$. Thus ${S^3}$ is ${S^2}$ worth of circles. Each circle, as a fiber ${S^1}$, threads through every other circle in the bundle ${S^3}$ without sharing a single point with any other circle, and projects down to a point such as $p$ on ${S^2}$ via the Hopf map ${h: S^3\rightarrow S^2}$, in a highly non-trivial configuration.}
\vspace{0.1cm}
\label{fig4}
\hrule
\end{figure*}

Geometrically the above results can be understood in terms of the twist in the Hopf fibration of ${S^3 \cong {\mathrm{SU}(2)}}$. Recall that locally (in the topological sense) ${S^3}$ can be written as a product ${S^2\times S^1}$, but globally it has no cross-section \cite{Ryder,Lyons}. It can be viewed also as a principal U(1) bundle over ${S^2}$, with the points of its base space ${S^2}$ being the elements of the Lie algebra su(2), which are pure quaternions, or bivectors \cite{disproof,Eguchi}. The product of two such bivectors are in general non-pure quaternions of the form (\ref{3}), and are elements of the group ${\mathrm{SU}(2)}$ itself. That is to say, they are points of the bundle space ${S^3}$, whose elements are the preimages\cite{Ryder} of the points of the base space ${S^2}$. These preimages are 1-spheres, ${S^1}$, called Hopf circles, or Clifford parallels \cite{Penrose}. Since these 1-spheres are the fibers of the bundle, they do not share a single point in common (cf. Fig.~\ref{fig4}). Each circle threads through every other circle in the bundle, making them linked together in a highly non-trivial configuration, which can be quantified by the following relation among the fibers \cite{Eguchi}:
\begin{equation}
e^{i\psi_-}\,=\,e^{i\phi}\,e^{i\psi_+}\,, \label{72-33}
\end{equation}
where ${e^{i\psi_-}}$ and ${e^{i\psi_+}}$, respectively, are the U(1) fiber coordinates above the two hemispheres ${H_-}$ and ${H_+}$ of the base space ${S^2}$, with spherical coordinates ${(0\leqslant\theta < \pi,\;0\leqslant\phi < 2\pi)}$; ${\phi}$ is the angle parameterizing a thin strip ${H_-\cap H_+}$ around the equator of ${S^2}$ [${\theta\sim\frac{\pi}{2}}$]; and ${e^{i\phi}}$ is the transition function that glues the two sections ${H_-}$ and ${H_+}$ together, thus constituting the 3-sphere. It is evident from Eq.${\,}$(\ref{72-33}) that the fibers match perfectly at the angle ${\phi=0}$ (modulo ${2\pi}$), but differ from each other at all intermediate angles ${\phi}$. For example,
${e^{i\psi_-}}$ and ${e^{i\psi_+}}$ differ by a minus sign at the angle ${\phi=\pi}$. Now in the Clifford-algebraic representation of our 3-sphere model the above relation can be written as
\begin{align}
\big\{\!\!-{\bf D}({\bf a})\,{\bf L}({\bf s}_1,\,\lambda^k)\big\}&\!=\!\big\{{\bf D}({\bf a})\,{\bf D}({\bf b})\big\}\big\{\!+{\bf L}({\bf s}_2,\,\lambda^k)\,{\bf D}({\bf b})\big\} \notag \\
&\Longleftrightarrow\;\;\;{\bf a}\,{\bf s}\,=\,\left\{{\bf a}\cdot{\bf b}+{\bf a}\wedge{\bf b}\right\}\;{\bf s}\,{\bf b}\,, \label{82}
\end{align}
provided we identify the angles ${\eta_{{{\bf a}}{{{\bf s}_1}}}}$ and ${\eta_{{{\bf s}_2}{{{\bf b}}}}}$ between ${{\bf a}}$ and ${{{\bf s}_1}}$ and ${{{\bf s}_2}}$ and ${{\bf b}}$ with the fibers ${\psi_-}$ and ${\psi_+\,}$, and the angle ${\eta_{{\bf a}{\bf b}}}$ between ${{\bf a}}$ and ${{{\bf b}}}$ with the generator of the transition function ${e^{i\phi}}$ on the equator of ${S^2}$. The above representation of Eq.(\ref{72-33}) is not as unusual as it may seem at first sight once we recall that geometric products of vectors and bivectors appearing in it are quaternions, and the quaternionic 3-sphere defined in Eq.${\,}$(\ref{onpara}) as a set of unit quaternions remains closed under multiplication. Indeed, as we saw in Eq.${\,}$(\ref{3}), each element of ${S^3}$ can be parameterized to take the form
\begin{align}
{\bf q}({\bf u},\,{\bf v})\,&=\,\cos(\,\eta_{{\bf u}{\bf v}})\,+\,\frac{{\bf u}\wedge{\bf v}}{||{\bf u}\wedge{\bf v}||}\,\sin(\,\eta_{{\bf u}{\bf v}}) \notag \\
&=\,\exp{\left\{\frac{{\bf u}\wedge{\bf v}}{||{\bf u}\wedge{\bf v}||}\;\eta_{{\bf u}{\bf v}}\right\}}\,, \label{defi-222}
\end{align}
which in turn can always be decomposed into a product of two bivectors, say ${{\boldsymbol\beta}({\bf u})}$ and ${{\boldsymbol\beta}({\bf v})}$, belonging to an ${S^2\subset S^3}$:
\begin{align}
-\,{\boldsymbol\beta}({\bf u})\,{\boldsymbol\beta}({\bf v})\,&=\,-\,(\lambda\,I\cdot{\bf u})\,(\lambda\,I\cdot{\bf v})  \,=\,{\bf u}\,{\bf v} \notag \\
&=\,\cos(\,\eta_{{\bf u}{\bf v}})\,+\,\frac{{\bf u}\wedge{\bf v}}{||{\bf u}\wedge{\bf v}||}\,\sin(\,\eta_{{\bf u}{\bf v}})\,. \label{defi-222-a}
\end{align}
Multiplying both sides of (\ref{82}) from the left with ${{\bf D}({\bf a})}$ and noting that all unit bivectors square to ${-1}$, we obtain 
\begin{equation}
{\bf L}({\bf s}_1,\,\lambda^k)\,=\,-\,{\bf D}({\bf b})\,{\bf L}({\bf s}_2,\,\lambda^k)\,{\bf D}({\bf b})\,.
\end{equation}
Next, multiplying the numerator and denominator on the RHS of this similarity relation with ${-{\bf D}({\bf b})}$ from the right and ${{\bf D}({\bf b})}$ from the left leads to the conservation of zero spin angular momentum, just as we have specified in Eq.${\,}$(\ref{56}):
\begin{align}
{\bf L}({\bf s}_1,\,\lambda^k)\,&=\,{\bf L}({\bf s}_2,\,\lambda^k) \notag \\
&\Longleftrightarrow\;\,
{\bf L}({\bf s}_{1},\,\lambda^k)\,{\bf L}({\bf s}_{2},\,\lambda^k)\,=\,{\bf L}^2({\bf s},\,\lambda^k)=\,-1, \label{596}
\end{align}
which was used in Eq.${\,}$(\ref{60}) to derive the strong correlations (\ref{abcos}). We have thus shown that the conservation of spin angular momentum is not an additional assumption, but follows from the very geometry and topology of the 3-sphere.

Returning to the Hopf fibration of ${S^3}$, it is not difficult to see from Eq.${\,}$(\ref{82}) that if we set ${{\bf a}={\bf b}}$ (or equivalently ${\eta_{{\bf a}{\bf b}}=0}$) for all fibers, then ${S^3}$ reduces to the trivial bundle ${S^2\times S^1}$, since then the fiber coordinates ${\eta_{{{\bf a}}{{{\bf s}_1}}}}$ and ${\eta_{{{\bf s}_2}{{{\bf b}}}}}$ would match up exactly on the equator of ${S^2}$ [${\theta\sim\frac{\pi}{2}}$]. In general, however, for ${{\bf a}\not={\bf b}}$, ${S^3\not=S^2\times S^1}$. For example, when ${{\bf a}=-\,{\bf b}}$ (or equivalently when ${\eta_{{\bf a}{\bf b}}=\pi}$) there will be a sign difference between the fibers at that point of the equator \cite{Ryder,Eguchi}. That in turn  would produce a twist in the bundle analogous to the twist in a M\"obius strip. It is this non-trivial twist in the ${S^3}$ bundle that is responsible for the observed sign flips in the product ${\mathscr{A}\mathscr{B}}$ of measurement events, from ${\mathscr{A}\mathscr{B}=-1}$ for ${{\bf a}={\bf b}}$ to ${\mathscr{A}\mathscr{B}=+1}$ for ${{\bf a}=-\,{\bf b}}$, as evident from the correlations (\ref{abcos}), which are obtained in the limits ${{\bf s}_1\rightarrow\,{\bf a}}$ and ${{\bf s}_2\rightarrow\,{\bf b}}$, together with ${{\bf s}_1={\bf s}_2={\bf s}}$, as in the definitions of the measurement functions (\ref{53}) and (\ref{54}). On the other hand, if the topology of our physical space were the trivial or product topology ${S^2\times S^1}$, then the transition function ${{\bf a}{\bf b}}$ in Eq.${\,}$(\ref{82}) would be identical to ${+1}$, and we would not observe sign flips from ${\mathscr{A}\mathscr{B}=-1}$ to ${\mathscr{A}\mathscr{B}=+1}$ when ${{\bf a}={\bf b}}$ is rotated to ${{\bf a}=-\,{\bf b}}$. Moreover, in that case the limits ${{\bf s}_1={\bf s}\rightarrow\,{\bf a}}$ and ${{\bf s}_2={\bf s}\rightarrow\,{\bf b}}$ would also reinforce ${{\bf a}{\bf b}=+1}$ in Eq.${\,}$(\ref{82}), which in turn would lead only to ${\mathscr{A}\mathscr{B}=-1}$ and never to ${\mathscr{A}\mathscr{B}=+1}$. Conversely, it is easy to see from the definitions (\ref{53}) and (\ref{54}) of ${\mathscr{A}}$ and ${\mathscr{B}}$ that, within the non-trivial topology of ${S^3}$ necessitated by the general transition function ${{\bf a}{\bf b}}$ in (\ref{82}), the relation ${\mathscr{A}= -\,\mathscr{B}}$ by itself does not impose any restrictions, such as ${{\bf a}={\bf b}}$, on the possible measurement directions ${\bf a}$ and ${\bf b}$ that Alice and Bob may wish to choose for their observations:
\begin{widetext}
\begin{align}
{\mathscr A}({\bf a},\,\lambda^k)
&\,=\,-\,{\mathscr B}({\bf b},\,\lambda^k) \\
\implies\;\;\;\lim_{{\bf s}\,\rightarrow\,{\bf a}}\left\{-\,{\bf D}({\bf a})\,{\bf L}({\bf s},\,\lambda^k)\right\}
&\,=\,-\,\lim_{{\bf s}\,\rightarrow\,{\bf b}}\left\{+\,{\bf L}({\bf s},\,\lambda^k)\,{\bf D}({\bf b})\right\} \\
\implies\;\;\;\lim_{{\bf s}\,\rightarrow\,{\bf a}}\left\{-\,\lambda^k\,{\bf L}({\bf a},\,\lambda^k)\,{\bf L}({\bf s},\,\lambda^k)\right\}
&\,=\,-\,\lim_{{\bf s}\,\rightarrow\,{\bf b}}\left\{+\,{\bf L}({\bf s},\,\lambda^k)\,\lambda^k\,{\bf L}({\bf b},\,\lambda^k)\right\} \\
\implies\;\;\;\lim_{{\bf s}\,\rightarrow\,{\bf a}}\left[\,-\lambda^k\left\{-\,{\bf a}\cdot{\bf s}\,-\,{\bf L}({\bf a}\times{\bf s},\,\lambda^k)\right\}\right]
&\,=\,-\,\lim_{{\bf s}\,\rightarrow\,{\bf b}}\left[\,+\lambda^k\left\{-\,{\bf s}\cdot{\bf b}\,-\,{\bf L}({\bf s}\times{\bf b},\,\lambda^k)\right\}\right] \\
\implies\;\;\;\lim_{{\bf s}\,\rightarrow\,{\bf a}}\left\{\,+\,\lambda^k\,{\bf a}\cdot{\bf s}\,+\,I\cdot({\bf a}\times{\bf s})\right\}
&\,=\,-\,\lim_{{\bf s}\,\rightarrow\,{\bf b}}\left\{\,-\,\lambda^k\,{\bf s}\cdot{\bf b}\,-\,I\cdot({\bf s}\times{\bf b})\right\} \\
\implies\;\;\;\lambda^k\,{\bf a}\cdot {\bf a}
&\,=\,\lambda^k\,{\bf b}\cdot {\bf b} \\
\implies\;\;\;||{\bf a}||^2
&\,=\,||{\bf b}||^2. \label{92}
\end{align}
\end{widetext}
This result dictates that only the unit magnitudes but not the directions of the vectors ${\bf a}$ and ${\bf b}$ are constrained to be equal, despite the apparent anti-correlation between ${\mathscr{A}}$ and ${\mathscr{B}}$ in their very definitions (\ref{53}) and (\ref{54}). Alice and Bob are thus free to choose any angle between ${\bf a}$ and ${\bf b}$ on the unit 2-sphere, in harmony with the fibration (\ref{82}) of ${S^3}$. The freedom to choose {\it any} directions ${\bf a}$ and ${\bf b}$ thus enables them to observe the twists in ${S^3}$, in the guise of the strong correlations (\ref{abcos}).

\section{Derivation of Tsirel'son's Bounds on the Strength of local-realistic Correlations}\label{apenB}

For completeness of our derivation of the correlation (\ref{65}), in this section we derive the Tsirel'son's bounds in (\ref{My-CHSH}) on the strength of such correlations. To this end, consider four observation axes, ${\bf a}$, ${\bf a'}$, ${\bf b}$, and ${\bf b'}$, for the experiment illustrated in Fig.~\ref{fig1}. Then the corresponding CHSH string of expectation values \cite{Christian,Bell-oversight}, namely the correlator
\begin{equation}
{\cal E}({\bf a},\,{\bf b})\,+\,{\cal E}({\bf a},\,{\bf b'})\,+\,
{\cal E}({\bf a'},\,{\bf b})\,-\,{\cal E}({\bf a'},\,{\bf b'})\,, \label{B1-11}
\end{equation}
would be bounded by the constant ${2\sqrt{2}}$, as discovered by Tsirel'son within the setting of Clifford algebra applied to quantum mechanics within a Hilbert space. Here each of the joint expectation values of the measurement results ${{\mathscr A}({\bf a},\,{\lambda})=\pm\,1}$ and ${{\mathscr B}({\bf b},\,{\lambda})=\pm\,1}$ are defined as
\begin{equation}
{\cal E}({\bf a},\,{\bf b})\,=\lim_{\,n\,\gg\,1}\left[\frac{1}{n}\sum_{k\,=\,1}^{n}\,{\mathscr A}({\bf a},\,{\lambda}^k)\;{\mathscr B}({\bf b},\,{\lambda}^k)\right],\label{exppeu}
\end{equation}
with the measurement functions ${{\mathscr A}({\bf a},\,{\lambda})}$ and ${{\mathscr B}({\bf b},\,{\lambda})}$ defined in (\ref{53}) and (\ref{54}). But from (\ref{57}) and (\ref{62}) we also have the following geometrical and statistical identity:
\begin{align}
\lim_{\,n\,\gg\,1}\Bigg[\frac{1}{n}\sum_{k\,=\,1}^{n}\,{\mathscr A}&({\bf a},\,{\lambda}^k)\;{\mathscr B}({\bf b},\,{\lambda}^k)\Bigg] \notag \\
&=\lim_{\,n\,\gg\,1}\left[\frac{1}{n}\sum_{k\,=\,1}^{n}\,{\bf L}({\bf a},\,\lambda^k)\,{\bf L}({\bf b},\,\lambda^k)\,\right].
\end{align}
Using this identity the correlator (\ref{B1-11}) can now be rewritten as the following single average:
\begin{align}
\lim_{\,n\,\gg\,1}\Bigg[\frac{1}{n}&\sum_{k\,=\,1}^{n}\big\{{\bf L}({\bf a},\,{\lambda}^k)\,{\bf L}({\bf b},\,{\lambda}^k)+{\bf L}({\bf a},\,{\lambda}^k)\,{\bf L}({\bf b'},\,{\lambda}^k) \notag \\
&+\,{\bf L}({\bf a'},\,{\lambda}^k)\,{\bf L}({\bf b},\,{\lambda}^k)+{\bf L}({\bf a'},\,{\lambda}^k)\,{\bf L}({\bf b'},\,{\lambda}^k)\big\}\Bigg]. \label{probnonint}
\end{align}
But since the bivectors ${{\bf L}({\bf a},\,\lambda)}$ and ${{\bf L}({\bf b},\,\lambda)}$ are two independent equatorial points of ${S^3}$, we can take them to belong to two disconnected ``sections'' of ${S^3}$
[{\it i.e.}, two disconnected su(2) 2-spheres within ${S^3\sim {\mathrm{SU}}(2)}$], satisfying
\begin{equation}
\left[\,{\bf L}({\bf n},\,{\lambda}),\,{\bf L}({\bf n'},\,{\lambda})\,\right]\,=\,0\,
\;\;\;\forall\;\,{\bf n}\;\,{\rm and}\;\,{\bf n'}\,\in\,{\mathrm{I\!R}}^3,\label{com}
\end{equation}
which is equivalent to anticipating a null outcome along the direction ${{\bf n}\times{\bf n'}}$ exclusive to both ${\bf n}$ and ${\bf n'}$. If we now square the integrand of equation (\ref{probnonint}), use the above commutation relations, and use the fact that all unit bivectors square to ${-1}$, then the absolute value of the Bell-CHSH string (\ref{B1-11}) leads to the following variance inequality \cite{Christian}:
\begin{align}
|{\cal E}&({\bf a},\,{\bf b})\,+\,{\cal E}({\bf a},\,{\bf b'})\,+\,
{\cal E}({\bf a'},\,{\bf b})\,-\,{\cal E}({\bf a'},\,{\bf b'})| \notag \\
&\leqslant\sqrt{\lim_{\,n\,\gg\,1}\left[\frac{1}{n}\sum_{k\,=\,1}^{n}\,
\big\{\,4\,+\,4\,{\mathscr T}_{\,{\bf a\,a'}}({\lambda}^k)\,{\mathscr T}_{\,{\bf b'\,b}}({\lambda}^k)\,\big\}\right]},\label{yever}
\end{align}
where the classical commutators
\begin{equation}
{\mathscr T}_{\,{\bf a\,a'}}(\lambda):=\frac{1}{2}\left[\,{\bf L}({\bf a},\,\lambda),\,{\bf L}({\bf a'},\,\lambda)\right]
\,=\,-\,{\bf L}({{\bf a}\times{\bf a'}},\,\lambda) \label{aa-potorsion-666}
\end{equation}
and
\begin{equation}
{\mathscr T}_{\,{\bf b'\,b}}(\lambda)
:=\frac{1}{2}\left[\,{\bf L}({\bf b'},\,\lambda),\,{\bf L}({\bf b},\,\lambda)\right]\,=\,-\,{\bf L}({{\bf b'}\times{\bf b}},\,\lambda)\label{bb-potor}
\end{equation}
are the geometric measures of the torsion within ${S^3}$. Thus it is the non-vanishing torsion ${\mathscr T}$ within the 3-sphere---{\it i.e.}, the parallelizing torsion which makes its Riemann curvature vanish---that is responsible for the stronger-than-linear correlation. We can see this from Eq.${\,}$(\ref{yever}) by setting ${{\mathscr T}=0}$, and in more detail as follows. By making a repeated use of the bivector identity
\begin{equation}
{\bf L}({\bf a},\,\lambda)\,{\bf L}({\bf a'},\,\lambda)\,=\,-\,{\bf a}\cdot{\bf a'}\,-\,{\bf L}({\bf a}\times{\bf a'},\,\lambda)\,,\label{bititi}
\end{equation}
the above inequality can be further simplified to
\begin{align}
&|{\cal E}({\bf a},\,{\bf b})\,+\,{\cal E}({\bf a},\,{\bf b'})\,+\,
{\cal E}({\bf a'},\,{\bf b})\,-\,{\cal E}({\bf a'},\,{\bf b'})| \notag \\
&\leqslant\sqrt{\!4-4({{\bf a}}\times{{\bf a}'})\cdot({{\bf b}'}\times{{\bf b}})-
4\!\lim_{\,n\,\gg\,1}\left[\frac{1}{n}\sum_{k\,=\,1}^{n}{{\bf L}}({\bf z},\,\lambda^k)\right]} \notag \\
&\leqslant\sqrt{\!4-4({{\bf a}}\times{\bf a'})\cdot({\bf b'}\times{{\bf b}})-
4\!\lim_{\,n\,\gg\,1}\left[\frac{1}{n}\sum_{k\,=\,1}^{n}\lambda^k\right]{{\bf D}}({\bf z})} \notag \\
&\leqslant\,2\,\sqrt{\,1-({{\bf a}}\times{\bf a'})
\cdot({\bf b'}\times{{\bf b}})\,-\,0\,}\,,\label{before-opppo-666}
\end{align}
where ${{\bf z}=({\bf a}\times{\bf a'})\times({\bf b'}\times{\bf b})}$. The last two steps follow from the relation (\ref{55}) between ${{\bf L}({\bf z},\,\lambda)}$ and ${{\bf D}({\bf z})}$ and the fact that the orientation ${\lambda}$ of ${S^3}$ is evenly balanced between ${+1}$ and ${-1}$. Finally, by noticing that trigonometry dictates
\begin{equation}
-1\leqslant\,({\bf a}\times{\bf a'})\cdot({\bf b'}\times{\bf b})\,\leqslant +1\,,
\end{equation}
the above inequality can be reduced to the familiar form
\begin{equation}
\left|\,{\cal E}({\bf a},\,{\bf b})\,+\,{\cal E}({\bf a},\,{\bf b'})\,+\,
{\cal E}({\bf a'},\,{\bf b})\,-\,{\cal E}({\bf a'},\,{\bf b'})\,\right|\,\leqslant\,2\sqrt{2}\,.
\label{B-CHSH}
\end{equation}
Needless to say, this result can also be derived directly from the correlation (\ref{65}):
\begin{align}
|\,{\cal E}({\bf a},&\,{\bf b})\,+\,{\cal E}({\bf a},\,{\bf b'})\,+\,{\cal E}({\bf a'},\,{\bf b})\,-\,{\cal E}({\bf a'},\,{\bf b'})\,| \notag \\
&=\,\left|\,-\,\cos\eta_{{\bf a}{\bf b}}\,-\,\cos\eta_{{\bf a}{\bf b'}}\,-\,\cos\eta_{{\bf a'}{\bf b}}\,+\,\cos\eta_{{\bf a'}{\bf b'}}\,\right| \notag \\
&\leqslant\,2\sqrt{2}\,.
\end{align}

\section{Standard interpretation of Bell's theorem is recovered within the flat geometry of ${{\mathrm{I\!R}^3}}$}\label{limit}

The ${S^3}$ model presented above becomes conducive to the traditional interpretation of Bell's theorem when the algebraic, geometrical and topological properties of the compactified physical space ${S^3}$ are ignored. In that case the upper bound of 2 on the Bell-CHSH inequality is respected. Thus, the results presented in this paper do not conflict with the standard interpretation of Bell's theorem outright but rather reproduces that interpretation as a special case in the flat geometry ${{\mathrm{I\!R}}^3}$ of the physical space, which is usually taken for granted in the literature on Bell's theorem. There are several different ways to appreciate this fact. As we saw in Section \ref{simu} above, one way to appreciate it is by analyzing the even-by-event simulations presented therein \cite{Simulation-A,Simulation-B}. Another way to appreciating it is by noting that if we ignore the twist (\ref{566}) or (\ref{596}) in the Hopf bundle of ${S^3}$, then the value of the correlation function ${{\cal E}({\bf a},\,{\bf b})}$ in (\ref{57}) reduces to ${-1}$ for all freely chosen parameters ${\bf a}$ and ${\bf b}$ for any initial state ${\lambda}$, and then the absolute bound of ${2}$ on the Bell-CHSH correlator (\ref{My-CHSH}) is not exceeded. A third way to appreciate it is by setting the torsion ${{\mathscr T}=0}$ in Eq.${\,}$(\ref{yever}) as noted between Eqs.~(\ref{bb-potor}) and (\ref{bititi}) during the derivation of the Tsirel'son's bounds in Section \ref{apenB}. Each of these three ways provide different insights into how the standard interpretation of Bell's theorem is recovered in ${{\mathrm{I\!R}}^3}$ limit.

\section{Theoretical and Experimental Support}

In recent years there has been some theoretical and experimental support for the local-realistic model of the strong correlations presented above. For example, in an influential recent paper published in {\it Nature Communications} \cite{Renner}, the authors state that ``Any no-go result, as for example Bell's theorem, is phrased within a particular framework that comes with a set of built-in assumptions. Hence it is always possible that a theory evades the conclusions of the no-go result by not fulfilling these implicit assumptions.'' This statement reflects a change in attitude of the physics community regarding the significance of Bell's theorem for fundamental physics. In Bell's local-realistic framework for the strong correlations \cite{Bell-1964}, there are built-in assumptions about the algebraic, geometrical and topological properties of the physical space in which we are confined to perform all our experiments, because it is modeled as ${\mathrm{I\!R}^3}$. In this paper we have removed these assumptions via greater rigor, by replacing ${\mathrm{I\!R}^3}$ with ${S^3}$ as the physical space. It is therefore not surprising that we have been able to reproduce the singlet correlations exactly. 

A more significant support for our locally causal model of the singlet correlations comes from a recently performed macroscopic experiment \cite{Macro-PRL}. It provides an important experimental confirmation of the model presented above. The authors of the experiment write: ``...~we have demonstrated the violation of a Bell-type inequality using massive (around ${10^{10}}$ atoms), macroscopic optomechanical devices, thereby verifying the nonclassicality of their state without the need for a quantum description of our experiment.'' To be sure, by nonclassicality the authors mean violation of local realism, and support this orthodox interpretation by providing a quantum mechanical description of their massive, macroscopic, mechanical system. However, the key phrase also used by the authors to describe their experiment is the following: ``...~without the need for a quantum description of our experiment.'' What this means is that we have an experimental proof that Bell-type inequalities can be violated also by purely classical, macroscopic systems without requiring a quantum mechanical description of the experiment. But that is exactly what our model presented above also predicts \cite{Christian, proposed}. According to our model, the violations of Bell inequalities are a result of the algebraic, geometrical, and topological properties of the compactified physical space, $S^3$. The concepts of quantum entanglement, non-locality, non-reality, or irreducible randomness are not necessary for explaining the violations of Bell inequalities.

\section{Concluding Remarks}

In this paper we have shown that it is possible to reproduce the statistical predictions of quantum mechanics in a locally causal manner, at least for the simplest entangled state such as the EPR-Bohm state. In particular, we have shown that such a locally causal description of the singlet state in the sense of Bell is possible at least within the spherical topology of a well known Friedmann-Robertson-Walker spacetime, viewed as a non-cosmological, terrestrial solution of Einstein's field equations. More specifically, we have presented a local, deterministic, and realistic model within such a Friedmann-Robertson-Walker spacetime which describes simultaneous measurements of the spins of two fermions emerging in a singlet state from the decay of a spinless boson. We have then shown that the predictions of this locally causal model agree exactly with those of quantum theory, without needing data rejection, remote contextuality, superdeterminism, or backward causation. A Clifford-algebraic representation of the 3-sphere with vanishing spatial curvature and non-vanishing torsion then allows us to transform our model in an elegant form. Several event-by-event numerical simulations of the model have confirmed our analytical results with accuracy of at least 4 parts in ${10^4}$.

\appendix 

\section{Formulation of Local Causality in the Manner of Bell}

\begin{figure*}[t]
\hrule
\vspace{0.2cm}
\scalebox{1.6}{
\begin{pspicture}(-0.7,-2.3)(-6.9,-4.95)

\begin{rotate}{-180}

\pscurve[linewidth=0.2mm,linestyle=dashed](3.25,2.64)(3.75,2.73)(5.05,3.6)(4.0,3.6)(2.85,2.75)(3.25,2.64)

\rput{180}(3.425,4.7){\scriptsize {${\bf v}$}}

\psline[linewidth=0.2mm,linestyle=dashed,arrowinset=0.3,arrowsize=2pt 3,arrowlength=2]{<-}(3.9,3.15)(3.5,4.5)

\pscurve[linewidth=0.2mm]{-}(3.65,3.25)(3.51,3.07)(4.3,3.25)(4.23,3.32)

\psline[linewidth=0.2mm,arrowinset=0.3,arrowsize=2pt 3,arrowlength=2]{->}(3.9,2.985)(4.0,3.005)

\rput{194}(4.3,2.65){\scriptsize {${I\cdot{\bf v}}$}}

\end{rotate}
\end{pspicture}}
\vspace{0.2cm}
\hrule
\caption{A unit bivector represents an equatorial point of a unit, parallelized 3-sphere. As shown in the figure, a bivector is an abstraction of a directed plane segment, with only a magnitude and a sense of rotation---{\it i.e.}, clockwise (${-}$) or counterclockwise (${+}$). Neither the depicted oval shape of its plane, nor its axis of rotation ${\bf v}$, is an intrinsic part of the bivector ${I\cdot{\bf v}}$.\break}
\label{fig5}
\hrule
\end{figure*}

In this appendix we review the notion of local causality, as originally conceived by Einstein in the present context, and later formalized by Bell \cite{Bell-1964}. A more detailed discussion by Bell on the subject can be found in his last paper \cite{Bell-1990}.

Our main goal here is to stress that the model constructed above is indeed locally causal in the sense of Einstein and Bell, despite the fact that it relies on the global topology of the spatial slices, ${S^3}$. It will also become evident from our discussion below that, although the correlation function ${{\cal E}({\bf a},\,{\bf b})}$ is manifestly time-independent, the measurement functions ${{\mathscr A}({\bf a},\,{\lambda})}$ and ${{\mathscr B}({\bf b},\,{\lambda})}$ it depends on are themselves not time-independent. Indeed they depend on the initial states ${\lambda}$ of the system specified at an earlier time and the final detector directions ${\bf a}$ and ${\bf b}$ chosen by Alice and Bob at a later time, as shown in Figs.${\,}$\ref{fig1} and \ref{fig2}. On the other hand, since we have set the scale factor ${a(t) = 1}$ in the solution (\ref{frw}), the times elapsed between the initial and final instants of the experiments are obviously not cosmological epochs.
For deterministic models of the EPR-Bohm correlation (such as the one constructed above), Bell considered a joint observable of the form ${{\mathscr A}{\mathscr B}({\bf a},\,{\bf b};\,\aleph,\,\lambda)=\pm1}$, where ${\bf a}$ and ${\bf b}$, respectively, are the freely chosen detector directions of Alice and Bob, ${\lambda}$ is an initial or ``complete'' state of the singlet system (which is also referred to as\footnote{Within the context of Bell's theorem ``shared randomness'' and ``hidden variable'' are used synonymously. See, for example, discussion in Ref.~\cite{Gisin}.} a ``common cause'', or ``shared randomness'', or ``hidden variable''), and ${\aleph}$ stands for any number of other pre-established constants and/or variables pertaining to the experimental set up, which we shall refer to as {\it shared background}. Here two of the most important differences between the variables ${\{{\bf a},\,{\bf b}\}}$ and the variables ${\{\aleph,\,\lambda\}}$ are: (1) while locally Alice and Bob have {\it total} control over the choice of variables ${\bf a}$ and ${\bf b}$ (respectively), they have {\it no} control over the variables ${\aleph}$ and ${\lambda}$ at any time during their experiment; and (2) while ${\aleph}$ and ${\lambda}$ are {\it completely specified} at an earlier time past the overlap of the backward light cones of Alice and Bob (cf. Fig.~\ref{fig2}), the variables ${\bf a}$ and ${\bf b}$ are freely chosen by them at a later time, as final directions along which the space-like separated measurement events ${{\mathscr A}({\bf a};\,\aleph,\,{\lambda})=\pm1}$ and ${{\mathscr B}({\bf b};\,\aleph,\,{\lambda})=\pm1}$ are determined. Bell called such events {\it locally explicable} if the joint observable ${{\mathscr A}{\mathscr B}({\bf a},\,{\bf b};\,\aleph,\,\lambda)=\pm1}$ of Alice and Bob can be factorized into local parts as
\begin{equation}
{\mathscr A}{\mathscr B}({\bf a},\,{\bf b};\,\aleph,\,\lambda)\,=\,{\mathscr A}({\bf a};\,\aleph,\,\lambda)\times{\mathscr B}({\bf b};\,\aleph,\,\lambda).
\end{equation}
Note that the functions ${{\mathscr A}({\bf a};\,\aleph,\,{\lambda})}$ and ${{\mathscr B}({\bf b};\,\aleph,\,{\lambda})}$ describe strictly local, realistic, and deterministically determined measurement events. Apart from the common cause ${\{\aleph,\,\lambda\}}$, which originates in the overlap of the backward light cones of Alice and Bob as shown in Fig.~\ref{fig2}, the event ${{\mathscr A}=\pm1}$ depends {\it only} on the measurement direction ${\bf a}$ chosen freely by Alice; and analogously, apart from the common cause ${\{\aleph,\,\lambda\}}$, the event ${{\mathscr B}=\pm1}$ depends {\it only} on the measurement direction ${\bf b}$ chosen freely by Bob. In particular, the function ${{\mathscr A}({\bf a};\,\aleph,\,\lambda)}$ {\it does not} depend on either ${\bf b}$ or ${\mathscr B}$, and the function ${{\mathscr B}({\bf b};\,\aleph,\,\lambda)}$ {\it does not} depend on either ${\bf a}$ or ${\mathscr A}$, just as demanded by Einstein's notion of local causality \cite{Bell-1964,Bell-1990}. The correlation between the simultaneous measurement results ${{\mathscr A}({\bf a};\,\aleph,\,\lambda)}$ and ${{\mathscr B}({\bf b};\,\aleph,\,\lambda)}$ can then be computed as
\begin{equation}
{\cal E}({\bf a},\,{\bf b})\,=\lim_{\,n\,\gg\,1}\left[\frac{1}{n}\sum_{k\,=\,1}^{n}\,
{\mathscr A}({\bf a},\,\aleph,\,{\lambda}^k)\;{\mathscr B}({\bf b},\,\aleph,\,{\lambda}^k)\right]\!. \label{prep}
\end{equation}

Now in the case of the local model constructed above the shared background ${\aleph}$ {\it includes} the topology ${\cal T}$ of the spatial slices ${S^3}$. And this topology is {\it completely specified} from the outset, past the overlap of the backward light cones of Alice and Bob. Therefore, Alice, for example, cannot influence either the freely chosen parameter ${\bf b}$, or the observed outcome ${\mathscr B}$ of Bob by altering the topology, say, from ${\cal T}$ to ${\cal T'}$. And likewise, Bob cannot influence either the freely chosen parameter ${\bf a}$, or the observed outcome ${\mathscr A}$ of Alice by altering the topology from ${\cal T}$ to ${\cal T'}$ ({\it e.g.}, from ${S^3}$ to ${{\mathrm{I\!R}}^3}$). Thus, despite its reliance on the global topology of spatial slices, there is no violation of local causality in our model.

It is also evident from the prescription (\ref{prep}) that, quite appropriately, the shared background ${\aleph}$ plays no role in the computation of the correlation. For this reason ${\aleph}$ is usually dropped from the measurement functions by writing them simply as ${{\mathscr A}({\bf a},\,{\lambda})}$ and ${{\mathscr B}({\bf b},\,{\lambda})}$, as we have done in this paper. On the other hand, from the above formulation of local causality it is evident that whether the joint outcome ${{\mathscr A}{\mathscr B}}$ is ${+1}$ or ${-1}$ depends on the elapsed time between the initial instant when the state ${\lambda}$ emerges from the source and the final instant when the measurements are made along the directions ${\bf a}$ and ${\bf b}$, within a spacetime specified by the Friedmann-Robertson-Walker solution (\ref{frw}), with ${a(t) = 1}$.

\section{Questions and Answers}

In no particular order, in this appendix we answer some questions concerning the local-realistic model for the singlet correlations presented in this paper.

\underbar{Question 1}: In the introduction it is claimed: "By contrast, in this paper we present a physically well-motivated constructive counterexample to Bell's theorem by deriving the strong singlet correlations using the powerful language of Geometric Algebra." This is extraordinary, because Bell's theorem is a mathematical theorem. It has been formally proven and experimentally tested. It is logically impossible to violate its conclusion without violating one of its premises. Yet, the paper is claiming to do just that. How is this possible?

\underbar{Answer 1}: Bell's theorem is {\it not} a theorem in the mathematical sense. It is a {\it physical} argument based on the mathematical inequalities discovered by George Boole some 100 years before they were used by Bell in his ``theorem.'' And, as a physical argument, Bell's theorem is a deeply flawed argument, as we have explained in Ref.~\cite{Bell-oversight} in considerable detail. In essence, as discussed also in the introductory section of this paper, ``contrary to the claim of Bell's theorem it is not the objectively measurable predictions of quantum mechanics that rule out the possibility of a local and realistic theory. It is the {\it ad hoc} and unjustified assumption of three or four physically incompatible experiments, any one of which might be performed on a given occasion, but only one of which can, in fact, be performed in practice, and in reality.'' In addition to this mistake, Bell's theorem is based on a number of other assumptions that can be and have been questioned.

\underbar{Question 2}: The paper wrongly suggests that Bell's framework assumes that ``physical space in which we are confined to perform all our experiments, is modeled as $\mathrm{I\!R}^3$." Bell's theorem contains no such assumption and holds independently of the space in which the ``hidden" variables are modeled.

\underbar{Answer 2}: Nowhere in his writings has Bell stated that his theorem holds independently of the physical space in which the hidden variables are modeled. In fact, Bell's proposed local-realistic framework {\it does} assume implicitly that physical space in which we are confined to perform all our experiments is modeled as $\mathrm{I\!R}^3$. It is unfortunate that this assumption is not made explicit by Bell and his followers in their writings. A deeper reflection on how the physical space is modeled in analyzing the Bell-test experiments is necessary to uncover this assumption. In the analyses of such experiments ordinary vector algebra (which does not, in fact, form an algebra) within $\mathrm{I\!R}^3$ is implicitly assumed. On the other hand, we have modeled the physical space as a quaternionic ${S^3}$ using Geometric Algebra. The fact that we have been able to reproduce the ``impossible'' strong correlations by modeling the physical space as quaternionic ${S^3}$ is a confirmatory evidence that the strategy to relax the implicit assumption built-in Bell's theorem has been successful.

\underbar{Question 3}: In Section \ref{secII} it is stated: ``Indeed, the pairs of measurement directions ${({\mathbf{a}},\,{\mathbf{b}})}$, ${({\mathbf{a}},\,{\mathbf{b'}})}$, ${({\mathbf{a'}},\,{\mathbf{b}})}$, and ${({\mathbf{a'}},\,{\mathbf{b'}})}$ are {\it mutually exclusive measurement directions}, corresponding to {\it incompatible} experiments which cannot be performed simultaneously.'' That is true, but in the proof of Bell's theorem measurements along these directions are assumed to be events in a probability space. The CHSH correlator (\ref{B1-11-2}) is therefore a random variable on that probability space. 

\underbar{Answer 3}: Unfortunately, no such joint probability space can meaningfully exist for the measurement events along the mutually exclusive pairs of directions ${({\mathbf{a}},\,{\mathbf{b}})}$, ${({\mathbf{a}},\,{\mathbf{b'}})}$, ${({\mathbf{a'}},\,{\mathbf{b}})}$, and ${({\mathbf{a'}},\,{\mathbf{b'}})}$. Therefore the claim that the CHSH correlator (\ref{B1-11-2}) is a random variable on such a space is not correct even mathematically. A joint probability space can exist only for compatible observables. But, inevitably, in the CHSH scenario what is involved for mutually exclusive directions are incompatible observables. Therefore the assumption of a joint probability space in such an argument is an invalid assumption. And even if we assume that we can perform Lebesgue integration over variables in such a fictitious probability space, it is possible to derive the Bell-CHSH inequality by simply considering four incompatible experiments without invoking the assumption of non-locality, as we have demonstrated in Ref.~\cite{Bell-oversight}. Consequently, any ``violation'' of the Bell-CHSH inequality is nothing but a consequence of the incompatibility of the four experiments.

\underbar{Question 4}: The violations of Bell inequalities, as they are demonstrated experimentally, are observed for a sequence of measurements, with the accumulation of statistics to determine the degree of violations, provided that the initial state is an appropriate entangled resource. The statement in Section \ref{secII} of the paper that ``We are therefore justified in ignoring the physical claim of Bell's theorem in this paper'' is therefore wrong. Bell's theorem may be viewed as a statement about how particular resource states (typically entangled states) violate the predictions of local realism. We may disagree about the derivation of Bell's theorem, or indeed how it pertains to observations. However, given that observations exist and appear to demonstrate convincing violations of Bell inequalities for appropriate non-classical resource states, we are not at liberty to ignore these established physical results.

\underbar{Answer 4}: Contrary to the frequently made claims such as the above, violations of the Bell inequalities is not what is demonstrated in the experiments at all. The sequence of measurements performed in the experiments, ``with the accumulation of statistics to determine the degree of violation'', are not theoretically bounded by 2 as claimed on the basis of Bell's theorem. They are, in fact, bounded by 4, and the bound of 4 is of course never exceeded in any experiment. Thus there is an extraordinary bait-and-switch happening (albeit unwittingly) in every experiment that claims to have violated the absolute bound of 2 on the Bell-CHSH inequality. We have explained this unwitting practice of bait-and-switch in more detail in Ref.~\cite{Bell-oversight} (which is also published as a section in Ref.~\cite{RSOS}). It is also important to note that predictions of the local-realistic ${S^3}$ model are identical to the predictions of quantum mechanics for the singlet state.

\underbar{Question 5}: In the proposed model, the measurement outcomes are supposed to be determined by one single vector ${\bf e}_o$; so the whole point of the contrived mathematical formalism the author has build up can be merely to define a probability distribution of ${\bf e}_o$.

\underbar{Answer 5}: This observation is not correct. It is evident from the definitions (\ref{aamset}) and (\ref{bamset}) of the measurement functions that they are determined by the detector directions ${\bf a}$ and ${\bf b}$, together with a {\it pair} of vectors ${({\bf e}_o,\,{\bf s}_o)}$, which form the initial state originating in the overlap of the backward light cones of Alice and Bob. As explained in the manuscript, all of the vectors involved in these definitions have specific geometrical meanings within a quaternionic 3-sphere, which we have used to represent the three-dimensional physical space. Moreover, there is nothing ``contrived'' about this geometrical representation, because it is a part of a well known solution of Einstein's field equations of general relativity. By contrast, defining a probability distribution of vector ${\bf e}_o$ without any relation to the geometry of the quaternionic 3-sphere and any reference to the vector ${\bf s}_o$ would not produce the model presented in this paper, and consequently it would not reproduce the strong correlations in a local-realistic manner.

\underbar{Question 6}: The statement below Eq.~(\ref{Cleq}) reads: ``Consequently, the detectors of Alice and Bob can receive the spin states ${{\bf e}_o}$ only if the constraints (\ref{rcon}) are satisfied.'' But surely the point of the experiment is that Alice and Bob receive particles every time, and the standard interpretation (derived ultimately from the Stern-Gerlach experiment) is that the measurement performs a projection onto the eigenstates of the measurement device. Hence it is not clear what is meant by Alice and Bob only receiving particular spin states.

\underbar{Answer 6}: The quoted sentence is from the middle of the paragraph that includes Eqs.~(\ref{Cleq}) and (\ref{cl48}). The preceding sentence reads: ``In our model the vectors ${{\bf e}_o}$ and ${{\bf s}_o}$ ensure in tandem that there are no initial states for which'' the constraint (\ref{Cleq}) is satisfied. This constraint, like those in (\ref{rcon}), arises from the geometry of the quaternionic 3-sphere, which, in the model, is taken as representing the physical space. The paragraph continues after Eq.~(\ref{cl48}) as follows: ``Clearly, a measurement event cannot occur if there does not exist a state which can bring about that event. Since the initial state of the system is specified by the pair ${({\bf e}_o,\,{\bf s}_o)}$ and not just by the vector ${{\bf e}_o}$, there are no states of the system for which ${|\cos(\,\eta_{{\bf n}{\bf e}_o})|\,<\,f(\eta_{{\bf z}{\bf s}_o})}$ for {\it any} vector ${\bf n}$. Thus a measurement event cannot occur for ${|\cos(\,\eta_{{\bf n}{\bf e}_o})|\,<\,f(\eta_{{\bf z}{\bf s}_o})}$, no matter what ${\bf n}$ is. As a result, there is a one-to-one correspondence between the initial state ${({\bf e}_o,\,{\bf s}_o)}$ selected from the set (\ref{lamset}) and the measurement events ${\mathscr A}$ and ${\mathscr B}$ specified by the Eqs.~(\ref{aamset}) and (\ref{bamset}).'' It is quite clear from this explanation that the spin states received by Alice and Bob are the only spin states that can exist within the quaternionic 3-sphere. What is more, the issue does not arise in the Geometric Algebra description of the model discussed in the later sections.

\underbar{Question 7}: Since matrix representation of the bivector subalgebra using Pauli matrices is equivalent to the bivector representation of the subalgebra within Geometric Algebra, why is the latter representation used rather than the former? 

\underbar{Answer 7}: The matrix representation of the bivector subalgebra fails at the very first step, because the product of two Pauli matrices can at most be an identity matrix, not a scalar number. But what are observed in the experiments, as results of the interactions between the spins ${{\bf L}({\bf a},\,\lambda)}$ and the detectors ${{\bf D}({\bf a})}$, are pure scalar numbers: ${{\mathscr A}({\bf a},\,{\lambda})=\pm\,1}$. Thus matrix representation is of no use in the present context. 

\underbar{Question 8}: In Eq.~(\ref{onpara}) the inner product ${I \cdot {\bf v}}$ between the trivector $I$ and the vector ${\bf v}$ is used, even though it would not be needed, since the outer product part of the geometric product ${I{\bf v}}$ is zero. Then why is the inner product ${I \cdot {\bf v}}$ used?

\underbar{Answer 8}: The inner product is used for a physical reason. We have the geometric product 
\begin{equation}
I{\bf v} = I\cdot{\bf v} + I \wedge {\bf v},
\end{equation}
which is equal to ${I\cdot{\bf v}}$ because, by definition, ${I \wedge {\bf v} \equiv 0}$ in the three-dimensional space. Thus, either ${I{\bf v}}$ or ${I\cdot{\bf v}}$ could have been used to represent the spin. But we have preferred to use ${I\cdot{\bf v}}$ instead of ${I{\bf v}}$ to represent spin because, up to sign, ${I\cdot{\bf v}}$ is identical to the dual of ${\bf v}$ (cf. Fig.~\ref{fig5}). ${\bf v}$ can thus be identified with the direction of measurement, freely chosen by the experimenters in any EPR-Bohm type experiment. Thus the choice of ${I\cdot{\bf v}}$ has been made for a physical reason.

\underbar{Question 9}: Eq.~(\ref{56}) requires ${\bf s}_1={\bf s}_2$. In Eq.~(\ref{59}) the two limits ${\bf s}_1\to {\bf a}$ and ${\bf s}_2\to {\bf b}$ are considered simultaneously, which makes sense only if ${{\bf a}={\bf b}}$. That makes the derivation (\ref{65}) of the singlet correlation ${\bf -{\bf a} \cdot {\bf b}}$ invalid for ${{\bf a}\not={\bf b}}$.

\underbar{Answer 9}: At the first sight the above argument may seem reasonable even though it contradicts the result (\ref{92}) where we have proved that simultaneous limits ${\mathbf{s}\to \mathbf{a}}$ and ${\mathbf{s}\to \mathbf{b}}$ do not necessitate the vector equality ${\mathbf{a}=\mathbf{b}}$, even if we incorrectly assume ${\mathscr{A}({\mathbf a},\,\lambda^k)=-\,\mathscr{B}({\mathbf b},\,\lambda^k)}$ for all measurement directions ${\mathbf{a}}$ and ${\mathbf{b}}$. There are very good physical and mathematical reasons for this. To begin with, the limits ${\mathbf{s}\to \mathbf{a}}$ and ${\mathbf{s}\to \mathbf{b}}$ are parts of the {\it independent} detection processes, captured in the measurement functions (\ref{53}) and (\ref{54}). These processes are {\it not} subject to the conservation law dictated by Eq.~(\ref{56}). They describe purely local interactions of the spin bivectors with the detector bivectors, occurring at spacelike separated observation stations of Alice and Bob. As we have illustrated in Fig.~\ref{fig1}, the singlet spin system ${-\,{\bf L}({\bf s}_1,\,\lambda^k)+{\bf L}({\bf s}_2,\,\lambda^k)}$ with vanishing total spin originates from the central source. Subsequently, the spin ${-{\bf L}({\bf s}_1,\,\lambda^k)}$ propagates freely towards Alice's detector ${{\bf D}({\bf a})}$ and the spin ${+{\bf L}({\bf s}_2,\,\lambda^k)}$ propagates freely towards Bob's detector ${{\bf D}({\bf b})}$. The conservation of zero spin angular momentum is maintained during this {\it free} evolution of the singlet system, giving rise to the equality ${\mathbf{s}_1=\mathbf{s}_2=\mathbf{s}}$ worked out in Eq.~(\ref{56}). But the vector ${\mathbf{s}_1}$ can remain equal to the vector ${\mathbf{s}_2}$ {\it only} until the start of the physical process of detection captured by the limit ${\mathbf{s}_1\to \mathbf{a}}$ encoded in Alice's measurement function ${\mathscr{A}({\mathbf a},\,\lambda^k)}$, and likewise for the vector ${\mathbf{s}_2}$ in the physical process of detection at Bob's end. This, as noted, is because the physical processes of detection at the two ends of the experiment are {\it not} subject to the conservation of zero spin. Therefore the initial impression that the limits ${\bf s}_1\to {\bf a}$ and ${\bf s}_2\to {\bf b}$ with ${\bf s}_1={\bf s}_2$ makes sense only if ${\bf a}={\bf b}$ is not correct. And, consequently, the derivation (\ref{65}) for $-{\bf a} \cdot {\bf b}$ is perfectly valid.

\underbar{Question 10}: The vectors and bivectors that enter the actual computation of the correlation (\ref{65}) are all space-like and use Clifford algebra of three-dimensional Euclidean space. The Clifford algebra of four-dimensional spacetime -- although discussed at the beginning of Section \ref{GA} -- does not enter the computation of the correlation (\ref{65}). Therefore Friedmann-Robertson-Walker spacetime is irrelevant for the ${S^3}$ model.

\underbar{Answer 10}: In the context of Bell's theorem the question of local causality is properly addressed only within a relativistic description of spacetime. See, for example, the discussion by Bell himself in his last paper on the subject \cite{Bell-1990}. In this paper Bell defines local causality in a given spacetime as follows:
\begin{quote}
A theory will be said to be locally causal if the probabilities attached to values of local beables in a space-time region 1 are unaltered by specification of values of local beables in a space-like separated region 2, when what happens in the backward light cone of 1 is already sufficiently specified, for example by a full specification of all local beables in a space-time region 3 (figure 6.4).
\end{quote}
Moreover, as is well known, a violation of the relativistic local causality can be separated into two conceptually distinct parts: (1) a signalling non-locality incompatible with special relativity, and (2) a non-signalling non-locality compatible with special relativity. These two conceptually distinct parts are kinematically captured by Bell in his definitions ${{\mathscr A}({\bf a},\,\lambda)}$ and ${{\mathscr B}({\bf b},\,\lambda)}$ of local measurement functions for any given initial state ${\lambda}$ of the system \cite{Bell-1964}. This separates relativistic local causality into independence of the parameter ${\bf a}$ from ${\bf b}$ (and vice versa) preserving signalling locality, and independence of the outcome ${\mathscr A}$ from ${\mathscr B}$ (and vice versa) preserving non-signalling locality. This separation allows one to recognize that quantum mechanics preserves {\it parameter independence} (thus remaining compatible with special relativity) but violates {\it outcome independence}. Thus, despite appearances, relativistic causality is implicit and essential in any discussion involving Bell-type measurement functions.

Now FRW spacetimes happen to be just the right spacetimes to be considered for addressing the question of local causality as we have done. That is because FRW spacetimes are sufficiently Newtonian to adequately host the correlations predicted by the non-relativistic singlet state (\ref{single}), and yet sufficiently relativistic to address the question of non-signalling non-locality that is suspected to be occurring at a spacelike distances in the EPR-Bohm type experiments. For this purpose, the condition of local causality to watch out for is that the initial state ${\lambda}$ that originates in the overlap of the backward light cones of Alice and Bob must bring about the measurement outcomes ${{\mathscr A}({\bf a},\,\lambda)}$ and ${{\mathscr B}({\bf b},\,\lambda)}$, for any freely chosen spacelike vectors ${\bf a}$ and ${\bf b}$. Thus it is not only the spacelike vectors and bivectors that play an essential role in understanding local causality within our model. The initial state ${\lambda}$ also plays its part to maintain relativistic causality in full four-dimensional spacetime picture, as dipicted in Fig.~\ref{fig2}.  
 
But what has been missing from the relativistic considerations by Bell in \cite{Bell-1990} are the algebraic, geometrical and topological properties of the physical space within which we are confined to perform all our experiments. And that is where the spacelike hypersurface, ${S^3}$, of a Friedmann-Robertson-Walker spacetime, enters our analysis. As explained in this paper, the geometry of the quaternionic 3-sphere is essential for the derivation of strong correlations, and that geometry is provided by the spacelike hypersurface of one of the three cosmological solutions of Einstein's field equations.

Suppose, however, we ignore the FRW line element (\ref{frw}) and start our analysis from Eq.~(\ref{onpara}) instead. But removing the FRW spacetime from the analysis in this manner would make the entire analysis {\it ad hoc}, with no physical justification for ${S^3}$. Thus, the claim that our analysis has nothing to do with Friedmann-Robertson-Walker spacetime is mistaken.

\underbar{Question 11}: In Eq.~(\ref{50}), {\it i.e.}, in the bivector subalgebra   
\begin{equation}
{\bf L}({\bf a},\,\lambda)\,{\bf L}({\bf b},\,\lambda)\,=\,-\,{\bf a}\cdot{\bf b}\,-\,{\bf L}({\bf a}\times{\bf b},\,\lambda)\,, \tag{\ref{50}}
\end{equation}
two different algebras are combined into the same equation. In other words, the bivectors appearing in the above identity are not all of the same kind, but a mixture of bivectors corresponding to two different algebraic representations.

\underbar{Answer 11}: This claim is not correct. Regardless of a given value of ${\lambda}$, $+1$ or $-1$, all three bivectors ${{\bf L}({\bf a},\,\lambda)}$, ${{\bf L}({\bf b},\,\lambda)}$, and ${{\bf L}({\bf a}\times{\bf b},\,\lambda)}$ in the above identity belong to the {\it same} algebraic representation of the standard bivector subalgebra (\ref{wh-o8899}). Thus, contrary to the claim, Eq.~(\ref{50}) does not describe two different multiplication rules but the same multiplication rule of the standard bivector subalgebra. The mistaken claim stems from a failure to understand what ${\lambda}$ stands for within ${S^3}$. It represents an orientation of the spin bivectors ${{\bf L}({\bf n},\,\lambda)}$ {\it relative} to the detector bivectors ${{\bf D}({\bf n})}$, as defined in Eq.~(\ref{55}). The meaning of ${\lambda}$ and the relationship between ${{\bf L}({\bf n},\,\lambda)}$ and ${{\bf D}({\bf n})}$ are clearly brought out between Eqs.~(\ref{pair1}) and (\ref{88}). They show that the left-handed subalgebra can be easily transformed into a right-handed subalgebra by reversing the order of the bivectors in their product, as verified also in the numerical simulations with a GAViewer program \cite{Wonnink,Diether-1,Diether-2}. Moreover, since ${\lambda}$ specifies the orientation of ${S^3}$ and not the handedness of a coordinate system [cf. Eq.~(\ref{un98})], the cross product ${{\bf a}\times{\bf b}}$ (which is of course universally defined by the right-hand rule) is not affected by it. The identity (\ref{50}) is simply a geometric product between the unit bivectors ${\mathbf{L}(\mathbf{a},\,\lambda)}$ and ${\mathbf{L}(\mathbf{b},\,\lambda)}$ representing the two spin angular momenta.

\underbar{Question 12}: In Eqs.~(\ref{57}) to (\ref{65}) two different representations of the bivector subalgebra are summed over illegally.

\underbar{Answer 12}: It is quite evident from these equations that what is being averaged over are the measurement results ${{\mathscr A}({\bf a},\,{\lambda})=\pm\,1}$ and ${{\mathscr B}({\bf b},\,{\lambda})=\pm\,1}$, which are limiting scalar points of a quaternionic 3-sphere as defined in the Eqs.~(\ref{53}) and (\ref{54}). Consequently, from Eqs.~(\ref{57}) and (\ref{62}) we have the following geometrical and statistical identity:
\begin{align}
&\lim_{\,n\,\gg\,1}\left[\frac{1}{n}\sum_{k\,=\,1}^{n}\,
{\mathscr A}({\bf a},\,{\lambda}^k)\;{\mathscr B}({\bf b},\,{\lambda}^k)\right] \notag \\
&\;\;\;\;\;\;\;\;\;\;\;\;\;\;\;\;\;\;\;\;\;=\lim_{\,n\,\gg\,1}\left[\frac{1}{n}\sum_{k\,=\,1}^{n}\,{\bf L}({\bf a},\,\lambda^k)\,{\bf L}({\bf b},\,\lambda^k)\,\right].
\end{align}
Evidently, all bivectors ${{\bf L}({\bf a},\,\lambda)}$ and ${{\bf L}({\bf b},\,\lambda)}$ in this identity belong to the {\it same} algebraic representation of the bivector subalgebra. In fact, the steps from (\ref{57}) to (\ref{62}) are quite straightforward and have been carefully explained just below Eq.~(\ref{65}). The steps from (\ref{62}) to (\ref{65}) are also straightforward. They follow at once upon using the relation (\ref{55}). While there is no room for a mistake in these latter three steps, they can be avoided by following Eqs.~(\ref{87}) to (\ref{stand-nossss}) instead, which provide an independent confirmation of the derivation from (\ref{57}) to (\ref{65}). Not surprisingly, both calculations give one and the same result (\ref{65}). What is more, two programmers have independently confirmed the validity of the derivation from (\ref{57}) to (\ref{65}) in two event-by-event numerical simulations of the singlet correlations using a GAViewer program \cite{Wonnink,Diether-1,Diether-2}.

\underbar{Question 13}: The correlations must be computed using actual experimental results such as ${{\mathscr A}({\bf a},\,{\lambda})=\pm\,1}$.

\underbar{Answer 13}: Yes, correlations must be computed using actual experimental results of ${+1}$ and ${-1}$, {\it but only to the extent that quantum mechanics is able to predict such actual measurement results}. After all, any local-realistic theory is obliged to reproduce only that which quantum mechanics is able to predict statistically and experimentalists are able to observe experimentally \cite{Clauser}. So, with that important correction to the claim, the correlations are indeed computed in the paper using actual experimental results of ${+1}$ and ${-1}$. Such actual experimental results are explicitly specified by the limiting scalar points ${{\mathscr A}({\bf a}\,,\,\lambda)=\pm1}$ and ${{\mathscr B}({\bf b}\,,\,\lambda)\pm1}$ of a quaternionic 3-sphere, which models the physical space in which we are confined to perform all our experiments. They correspond exactly to the measurement results considered by Bell in his paper (cf. Eq.~(1) of Ref.~\cite{Bell-1964} and Eqs.~(\ref{53}) and (\ref{54}) of this paper). These ${+1}$ or ${-1}$ results are then averaged over in Eq.~(\ref{57}), which is {\it the} standard way of computing the correlations in the experimental context of Bell's theorem.

\underbar{Question 14}: According to the definitions (\ref{53}) and (\ref{54}) we can identify ${{\mathscr A}({\bf a},\,\lambda)}$ with ${-\lambda}$ and ${{\mathscr B}({\bf b},\,\lambda)}$ with ${+\lambda}$ so that ${{\mathscr A}({\bf a},\,\lambda){\mathscr B}({\bf b},\,\lambda)=(-\lambda)(+\lambda)=-1}$ for all ${\bf a}$ and ${\bf b}$, which immediately gives the correlation ${{\cal E}({\bf a},\,{\bf b})=-1}$ for all ${\bf a}$ and ${\bf b}$ even when ${{\bf b}\not={\bf a}}$, and that contradicts the result (\ref{65}).

\underbar{Answer 14}: There are a number of physical reasons why such identification of ${{\mathscr A}({\bf a},\,\lambda)}$ with ${-\lambda}$ and ${{\mathscr B}({\bf b},\,\lambda)}$ with ${+\lambda}$ is wrong. To begin with, it confuses the measurement outcomes ${{\mathscr A}=\pm1}$ and ${{\mathscr B}=\pm1}$ (which are observed by Alice and Bob at spacelike separated stations) with the initial state ${\lambda=\pm1}$ (which originates at the central source in the overlap of the backward light cones of Alice and Bob, as shown in Fig.~\ref{fig2}). But it is evident from the definitions (\ref{53}) and (\ref{54}) that the measurement results ${{\mathscr A}({\bf a},\,\lambda)}$ and ${{\mathscr B}({\bf b},\,\lambda)}$ come about as a consequence of interactions between the detector bivectors and spin bivectors, with the latter originating at the source. Moreover, it is easy to recognize from the definitions (\ref{53}) and (\ref{54}) and Eq.~(\ref{59}) that the product of the functions ${{\mathscr A}({\bf a},\,\lambda)}$ and ${{\mathscr B}({\bf b},\,\lambda)}$ is in general a unit quaternion that is not equal to $-1$. It may take both values, $+1$ and $-1$, in the scalar limits. This is evident from Eqs.~(\ref{56}) and (\ref{566}), which dictate that ${{\mathscr A}{\mathscr B}=-1}$ for ${{\bf b}\not={\bf a}}$ can occur if and only if the conservation of spin angular momentum is violated. Finally, our goal is to recognize that ${{\mathscr A}=\pm1}$ and ${{\mathscr B}=\pm1}$ are limiting scalar points of a quaternionic 3-sphere and the correlation between them is ${{\cal E}({\bf a},\,{\bf b})=-{\bf a}\cdot{\bf b}}$. Identification of ${{\mathscr A}({\bf a},\,\lambda)}$ with ${-\lambda}$ and ${{\mathscr B}({\bf b},\,\lambda)}$ with ${+\lambda}$ frustrates that goal. On the other hand, if we ignore the twist (\ref{566}) or (\ref{596}) in the Hopf bundle of ${S^3}$ (which, as we saw in Section \ref{raison}, would be equivalent to violating the conservation of zero spin angular momentum), then the value of the correlation function ${{\cal E}({\bf a},\,{\bf b})}$ given in (\ref{57}) reduces to ${-1}$ for all directions  ${\bf a}$ and ${\bf b}$, for any initial state ${\lambda}$, and then the absolute bound of ${2}$ on the Bell-CHSH correlator (\ref{B1-11-2}) is not exceeded.

\underbar{Question 15}: The 3-sphere model appears to be conspiratorial and thus uninteresting. The initial state $\lambda$ or its probability distribution seems to depend on the detector orientation, in violation of ``no conspiracy'' assumption of Bell's theorem.

\underbar{Answer 15}: The claim that the initial state or its probability distribution depends on the detector orientation is based on a misunderstanding of what is meant by ``orientation'' in Geometric Algebra. An ``orientation'' in Geometric Algebra means {\it handedness}. For the model to be conspiratorial in the sense of Bell, the hidden variable $\lambda$ would have to depend on the detector settings $\mathbf{n}$, or vice versa. But that is not the case in the model presented in the paper. Alice and Bob are completely free to choose the settings $\mathbf{n}$ appearing in the detectors ${\mathbf{D}(\mathbf{n})}$, independently of the value of $\lambda$, and vice versa. ${\mathbf{D}(\mathbf{n})=I\cdot\mathbf{n}}$ is a unit bivector, which is literally equal to ``$+1$ about the vector $\mathbf{n}$'', where $\mathbf{n}$ is a {\it freely chosen} experimental parameter. It is thus abundantly clear that $\mathbf{n}$ does not depend on $\lambda$, and $\lambda$ does not depend on $\mathbf{n}$. Consequently, there is no conspiracy in the model in the sense of Bell.

\underbar{Question 16}: How can the results presented in this paper be used for quantum computing problems or the foundations of quantum mechanics without their radical modification?

\underbar{Answer 16}: A significant portion of the quantum computing enterprise relies on the concept of quantum entanglement as a fundamental feature of the world. However, what can be actually observed in any physical experiment is not quantum entanglement but the strong correlations predicted by the entangled quantum system. In the manuscript we have shown that the strong correlations predicted by the singlet state (\ref{single}) can be explained without the concept of entanglement and the associated notion of irreducible randomness. They can be explained as local, realistic, and deterministic correlations among the points of a quaternionic 3-sphere without the need for the notion of irreducible randomness, which is at the heart of the effort to create scalable quantum computers. It is, however, beyond the scope of the present manuscript to demonstrate in detail how the results presented therein can be used for solving the quantum computing problems.

\underbar{Question 17}: How are the predictions of the local model described using density matrices instead of state vectors?

\underbar{Answer 17}: The singlet state ${|\Psi_{\bf n}\rangle}$ expressed in Eq.~(\ref{single}) and the EPR-Bohm correlations it predicts can be described also using a density matrix. However, the corresponding local-realistic computation of the correlations in terms of algebra, geometry and topology of the quaternionic 3-sphere would be exactly the same. This is because the quantum mechanical correlations for the joint spin observable ${{\boldsymbol\sigma}_1\cdot{\bf a}\otimes{\boldsymbol\sigma}_2\cdot{\bf b}}$ can be computed by either using the state vector ${|\Psi_{\bf n}\rangle}$ or using the density matrix ${\cal W}$, as follows: 
\begin{align}
{\cal E}_{q.m.}({\bf a},\,{\bf b})\,&=\,
\langle\Psi_{\bf n}|\,{\boldsymbol\sigma}_1\cdot{\bf a}\,\otimes\,
{\boldsymbol\sigma}_2\cdot{\bf b}\,|\Psi_{\bf n}\rangle\, \notag \\
&=\,\text{Tr}\left\{{\cal W}\left({\boldsymbol\sigma}_1\cdot{\bf a}\,\otimes\,
{\boldsymbol\sigma}_2\cdot{\bf b}\right)\right\} \notag \\ 
&=\,-\,
{\bf a}\cdot{\bf b}\,.
\end{align}
Since the goal in the manuscript is to reproduce only the correlations in a local-realistic manner, it is irrelevant how they have been computed using a quantum mechanical method ({\it i.e.}, whether using a state vector or a density matrix).

\acknowledgments

I am grateful to Fred Diether, Michel Fodje, Chantal Roth, and Albert Jan Wonnink for discussions and help with the numerical simulations, and to Jay R. Yablon for the result (\ref{92}), and for refuting criticism of the derivation of Eq.~(\ref{600}). The simulation \cite{Simulation-A} was inspired by the original simulation by Michel Fodje \cite{Fodje}. The distribution function used by Fodje was improved by Richard D. Gill in 2014 with help from me and several other participants in an online discussion forum, leading to the distribution function (\ref{ill}) (see also references cited in the simulation \cite{Simulation-A}). The two distribution functions produce nearly identical correlations. The first version \cite{Local} of this paper was written in 2014 at the Wolfson College of Oxford University, and published as a preprint on the arXiv. I have replied to online criticisms of the first version in \cite{Aaronson}.

\end{document}